\documentclass[a4paper,fleqn,usenatbib]{mnras}

\usepackage{mathptmx}

\usepackage[T1]{fontenc}
\usepackage{ae,aecompl}

\usepackage{graphicx}	
\usepackage{amsmath}	
\usepackage{amssymb}	

\usepackage{multicol}        
\usepackage{bm}         
\usepackage{pdflscape}  

\usepackage{enumitem}
\usepackage{xspace}
\usepackage{hhline}

\newcommand{\ltsima}{$\; \buildrel < \over \sim \;$}
\newcommand{\lsim}{\lower.5ex\hbox{\ltsima}}
\newcommand{\gtsima}{$\; \buildrel > \over \sim \;$}
\newcommand{\gsim}{\lower.5ex\hbox{\gtsima}}

\newcommand{\swift} {{\it Swift}\xspace}
\newcommand{\swiftxrt} {{\it Swift}/XRT\xspace}
\newcommand{\swiftuvot} {{\it Swift}/UVOT\xspace}
\newcommand{\swiftbat} {{\it Swift}/BAT\xspace}
\newcommand{\swiftxrtuvot} {{\it Swift}/XRT and UVOT\xspace}
\newcommand{\maxi} {{\it MAXI}\xspace}
\newcommand{\rxteasm} {{\it RXTE}/ASM\xspace}
\newcommand{\suzaku} {{\it Suzaku}\xspace}
\newcommand{\smarts} {{\it SMARTS}\xspace}
\newcommand{\azt} {{\it AZT33IK}\xspace}
\newcommand{\rtt} {{\it RTT150}\xspace}
\newcommand{\zeiss} {{\it ZEISS1000}\xspace}

\newcommand{\tubitak} {{\it T{\"U}B{\.I}TAK}\xspace}

  \title[Broad-band SED evolution during outburst rise in Aql~X-1]{Evolution of broad-band SED during outburst rise in NS X-ray Nova Aql~X-1}
  \author[Meshcheryakov~et~al.]{Alexander~V.~Meshcheryakov$^{1,2}$\thanks{E-mail:~mesch@iki.rssi.ru},
    Sergey S. Tsygankov$^{3}$, Irek M. Khamitov$^{2,4}$, \newauthor
  Nikolay I. Shakura$^5$, Ilfan F. Bikmaev$^{2,6}$, Maxim V. Eselevich$^{7}$, 
    Valeriy V. Vlasyuk$^{8}$
    \\
    $^{1}$ Space Research Institute of the Russian Academy of Sciences
    (IKI), 84/32 Profsoyuznaya Str, Moscow, Russia, 117997 \\
    $^{2}$ Kazan Federal University, Kremlevskaya 18, 420008 Kazan, Russia \\
    $^{3}$ Tuorla Observatory, Department of Physics and Astronomy, University of Turku,
  V\"ais\"al\"antie 20, FI-21500 Piikki\"o, Finland \\
    $^{4}$ \tubitak National Observatory (TUG), 07058, Akdeniz University Campus, Antalya, Turkey \\
    $^{5}$ Sternberg Astronomical Institute, Moscow M. V. Lomonosov State University, Universitetskij pr., 13, Moscow 119992, Russia\\
    $^{6}$ Academy of Sciences of Tatarstan, Bauman str., 20, Kazan, Russia\\
    $^{7}$ Institute of Solar-Terrestrial Physics of Siberian Brunch of Russian Academy of Sciences (ISTP SB RAS), Irkutsk, Russia \\
    $^{8}$ Special Astrophysical Observatory of Russian Academy of Sciences (SAO RAS), 369167, Nizhnij Arkhyz,  Russia }

\begin{document}

\date{}

\pagerange{\pageref{firstpage}--\pageref{lastpage}} \pubyear{2017}
\date{Accepted XXX. Received YYY; in original form ZZZ}

\maketitle

\label{firstpage}

\begin{abstract}
The observed evolution of the broad-band spectral energy distribution (SED) in NS X-ray Nova Aql~X-1 during the rise phase of a bright FRED-type outburst in 2013 can be understood in the framework of thermal emission from unstationary accretion disc with temperature radial distribution transforming from a single-temperature blackbody emitting ring into the multi-colour irradiated accretion disc. SED evolution during the hard to soft X-ray state transition looks curious, as it can not be reproduced by the standard disc irradiation model with a single irradiation parameter for NUV, Optical and NIR spectral bands. NIR (NUV) band is correlated with soft (hard) X-ray flux changes during the state transition interval, respectively. In our interpretation, at the moment of X-ray state transition UV-emitting parts of the accretion disc are screened from direct X-ray illumination from the central source and are heated primary by hard X-rays ($E>10$\,keV), scattered in the hot corona or wind possibly formed above the optically-thick outer accretion flow; the outer edge of multi-colour disc, which emits in Optical-NIR, can be heated primary by direct X-ray illumination.     

We point out that future simultaneous multi-wavelength observations of X-ray Nova systems during the fast X-ray state transition interval  are of great importance, as it can serve as "X-ray tomograph" to study physical conditions in outer regions of accretion flow. This can provide an effective tool to directly test the energy-dependent X-ray heating efficiency, vertical structure and accretion flow geometry in transient LMXBs.  
\end{abstract}


\section{Introduction} 
X-ray Novae (XN), also called as Soft X-ray Transients (SXT), are Low Mass X-ray Binaries (LMXB) showing transient accretion activity. During the accretion outburst the luminosity of the system in the X-ray spectral range, where the main energy release happens, rises up to $10^{6}$ times with respect to quiescence level. The observational studies of X-ray Nova systems are on fundamental importance for the physics of extreme states of matter. The majority ($\sim75$\%) of X-ray Novae systems contain a black hole candidate as a primary star \citep{2016AA...587A..61C}. 


Despite many existing studies of multi-wavelength light curves of outbursts in various transient LMXBs (see e.g. \cite{1997ApJ...491..312C,2000ApJ...532.1069E,2008ApJ...688..537M,2009MNRAS.392.1106G, 2014ApJ...784..122D,2014PASJ...66...84N,2016AstL...42...69G} and many other studies) definitely there is a lack of a detailed analysis focused on the beginning parts of XN outbursts (covering the stage of initial flux rise from the quiescent state to the outburst maximum).  A substantial attention is payed to the analysis of decaying parts of FRED-type events (see e.g. \cite{2008AA...491..267S,2016arXiv161001399L}), which can be well reproduced in theoretical models of XN outbursts \citep{2001AA...373..251D}. The outburst rise phase in X-ray Novae is much less studied, due to a lack of a good quality multi-wavelength observational data during this time period. The fast rise stage in X-ray Novae has usually a much poorer coverage by multi-wavelength observations, mainly because of relatively late detection of a new outburst by currently on-orbit X-ray monitors (e.g. MAXI, {\it SWIFT}/BAT). The existing studies covering the outburst rise period in X-ray Novae are concentrated primary on measurement and interpretation of possible time delays between IR-Optical-UV and X-ray light curves (see e.g. \cite{1997ApJ...489..234H,1998MNRAS.300.1035S,2016arXiv160507013B}). For the development of a better model of accretion flow during XN outbursts, it is important to compare the spectral evolution predicted by the common theory of unstationary disc accretion and the observed  spectral energy distribution (SED) evolution during outburst rise phase in real X-ray Nova systems.

In this work, we perform a detailed study of the broad-band SED evolution during the outburst rise phase in the famous NS X-ray Nova system Aql~X-1 --- the most prolific SXT known to-date. We present a multi-wavelength observational data for the initial rising phase of bright outburst in 2013, carried out during the monitoring campaign of Aql~X-1 at Swift orbital observatory and a few 1-m class ground-based optical telescopes. Our main aim here is to qualitatively compare the observed broad-band SED evolution in this prototypical NS X-ray Nova system to theoretical expectations for the model of non-stationary accretion disc, which is developed during the outburst rise phase.  

The article is organised as follows. In the \S\ref{sec:aqlx1} we describe Aql~X-1 system, its orbital and accretion disc parameters and interstellar extinction to the source. In \S\ref{sec:reduction} our observational data and its reduction are described. In the \S\ref{sec:results} we present a multi-band light curves and derived SED measurements for the rising part of Aql~X-1 outburst, as well as adopted spectral models. In the section \S\ref{sec:xray_tomograph} we discuss the  "X-ray tomograph" effect, working at the moment of X-ray state transition in Aql~X-1, as a promising observational tool for direct testing of the energy-dependent X-ray heating efficiency and vertical structure of accretion disc in X-ray Novae systems. Our results for Aql~X-1 broad-band SED evolution during outburst rise phase are presented in the section \S\ref{sec:analysis_discussion}. In the last section our conclusions are drawn.    

\section{Aql~X-1}
\label{sec:aqlx1}

 Aql X-1 is a transient X-ray binary system in which a compact object accretes matter from an accretion disc which is supplied by the Roche lobe filling low mass companion. With more than 40 outbursts observed in the X--ray and/or optical bands since its discovery in 1965 \citep{1967Sci...156..374F}, Aql X-1 is the most prolific X--ray transient known to date (about 25 outbursts were detected in the 1996--2016 epoch). Observations of type I X--ray bursts \citep{1981ApJ...247L..27K} and coherent millisecond X-ray pulsations \citep{2008ApJ...674L..41C,2016arXiv161102578T} lead to a surely identification of the compact object in this system as a neutron star. Aql~X-1 X-ray spectral and timing behaviour classify it as an atoll source \citep{2000ApJ...530..916R}.

The optical counterpart of Aql~X-1 is known to be an evolved K4$\pm$2 spectral type star \citep{2016arXiv160900392M}, with a quiescent magnitude of $21.6\pm0.1$ mag in the $V$ band \citep{1999AA...347L..51C}. An interloper star located only $0.48''$ east of the true counterpart heavily complicates the studies in the quiescent state (\cite{1999AA...347L..51C}; \cite{2012ApJ...749....3H}). In the recent high angular resolution near infrared spectroscopy observations \citep{2016arXiv160900392M} the first dynamical solution for Aql~X-1 was obtained. 

Despite its frequent outbursts, there are few reported radio detections  of  Aql  X-1,  likely  owing  to  the  faintness  of  atoll sources  in  the  radio  band \citep{2006MNRAS.366...79M}. The available observations suggest that the radio emission is being activated by both transitions from a hard state to a soft state and by the reverse transition at lower X-ray luminosity. The maximum radio flux density $0.68^{\pm0.09}$\,mJy (8.4GHz) was detected at the moment of state transition during Aql~X-1 outburst in Nov 2009 \citep{2010ApJ...716L.109M}. In all available multi-wavelength observations, the radio spectrum was flat or inverted, with flux density scaling as $F_\nu\propto\nu^{\gtrsim0}$ \citep{2009MNRAS.400.2111T}. There is evidence for quenching of the radio emission at X-ray fluxes above $5\cdot10^{-9}$\,erg/s/cm$^2$ ($L_{X}\gtrsim0.1{}L_{\rm Edd}$) \citep{2010ApJ...716L.109M}.

\begin{table}
\caption{Aql~X-1 system parameters}
\begin{tabular}{lll}
Parameter & Value & Reference \\
\hline
$P_{orb}$ [d]      & $0.7895138(10)$         & \cite{2016arXiv160900392M} \\
$T_{0}$ [MJD]      & $55809.895(5)$     & \cite{2016arXiv160900392M} \\
$i$                & $42\pm4^\circ$          & \cite{2016arXiv160900392M} \\
$m_1$ [$M_\odot$]  & $>1.2\approx1.4$        & \cite{2013ApJ...778...66K}, \\
                                                          & & \cite{1981ApJ...247L..27K} \\
$q$                & $0.39\pm0.14$  & \cite{2016arXiv160900392M} \\
$D$ [kpc]          & $5.0\pm0.9$ & \cite{2008ApJS..179..360G} \\
$N_H$ [$cm^{-2}$]  & $3.6\cdot10^{21}$   & see \S\ref{sec:aqlx1}\\
$E_{B-V}$[$mag$]   & $0.65$             & see \S\ref{sec:aqlx1}\\
\hline
\end{tabular}
\label{tbl:aqlx1_pars} 
\end{table}

\paragraph*{Orbital and accretion disc parameters of Aql~X-1.}
Orbital parameters of Aql~X-1 are well defined by previous extensive observational studies of this X-ray Nova system. In the Table~\ref{tbl:aqlx1_pars} we provide a best estimates for system parameters (orbital period $P_{orb}$, primary mass in solar units $m_1$ , mass ratio $q=m_2/m_1$, system inclination $i$), distance to the source $D$ and ephemeris for the time of the minimum of the outburst light curve $T_0$ (phase zero corresponds to inferior conjunction of the secondary star), which we will use throughout this paper. 

By using the Aql~X-1 binary system orbital parameters, we derived the characteristic accretion disc radii and the Roche lobe sizes for the primary and secondary star in the binary system, in the following way. First, with reasonable assumption of LMXB eccentricity $e=0$, the major semi-axis in the binary system $a$ can be estimated from the Kepler's law as:
\begin{equation} 
  \label{eq:a_Kepl}
  a = 3.52\times10^{10} m_1^{1/3}(1+q)^{1/3}  
  \Bigg(\frac{P_{orb}}{1 h}\bigg)^{2/3} \approx 3.1\times10^{11}\,cm ~.
\end{equation}
The effective radius of the Roche lobe of the primary ($R_{L1}$) and the secondary ($R_{L2}$) stars in the close binary system can be obtained from \citep{1983ApJ...268..368E}. The compact object Roche lobe radius in Aql~X-1 system was estimated as
\begin{equation}
\label{eq:RL1}
\frac{R_{L1}}{a}= \frac{0.49}{0.6 + q^{2/3} ln(1+q^{-1/3})}\approx 0.46 ~. 
\end{equation}
For $R_{L2}$, one need to replace $q\rightarrow{}q^{-1}$ in the formula above:
\begin{equation}
\label{eq:RL1}
\frac{R_{L2}}{a} \approx 0.30 ~. 
\end{equation}
Due to the angular momentum conservation of accreting matter, the disk radius can not be smaller than the circularisation radius (see numerical simulation in \cite{1975ApJ...198..383L} and their analytic approximation in \cite{2005astro.ph..1215G}, $3$\% accurate for $0.03\le{}q\le{}10$): 
\begin{equation}
\frac{R_{\rm circ}}{a}=0.074\ \left ( \frac{1+q}{q^2} \right )^{0.24}\approx 0.13 ~. 
\end{equation}
The maximal outer radius of the accretion disc in LMXB can be estimated by the tidal truncation radius (see numerical simulation in \cite{1977ApJ...216..822P} and their analytic approximation in \citep{2005astro.ph..1215G}, 3\% accurate for $0.06<q<10$ range):  
\begin{equation}
  \frac{R_{\rm tid}}{a} = 0.112 + \frac{0.270}{1+q} + \frac{0.239}{(1+q)^2} \approx 0.43 ~.
  \label{eq:Rtidal}
\end{equation} 
     
\paragraph*{Extinction to Aql~X-1 in X-rays and NUV-NIR.}
Extinction in the X-ray spectral range in the direction to Galactic LMXBs is caused by photoionisation effect in the interstellar gas on the line-of-sight (if internal extinction in the vicinity of the source is negligible). With reasonable assumption of solar chemical abundance, value of X-ray extinction to the source depends only on the hydrogen column density ($N_H$) parameter. Extinction in the NUV-NIR spectral range in the Galaxy is caused by absorption on the interstellar dust grains. We adopted standard extinction law \citep{1989ApJ...345..245C} with fixed $R_V=3.1$. Then value of NUV-NIR extinction depends only on color excess ($E_{B-V}$) parameter. Below in this section we obtain best estimates for $N_H$ and $E_{B-V}$ for Aql~X-1. 

First, we estimated maximum $N_H$ in the direction to Aql~X-1 by using the common Leiden/Argentine/Bonn (LAB) Survey of Galactic HI \citep{2005AA...440..775K} and HI map of \citep{1990ARAA..28..215D} (DL). nH routine from FTOOLS library (\cite{1995ASPC...77..367B}, http://heasarc.gsfc.nasa.gov/ftools/) was used. Note that these maps have limited resolution of approximately 0.5 degrees and 1 degrees, respectively. We obtained the following estimates for hydrogen column density: $n_H\approx2.48\times10^{21}$~atoms$\cdot$cm$^{-2}$ (LAB) and $n_H\approx3.43\times10^{21}$~atoms$\cdot$cm$^{-2}$ (DL). As we see the substantial dispersion between two estimates and taking into account the possibility of internal extinction in the source, we decided to adopt as best $N_H$ the value obtained from spectral fits during the outburst stage in Aql~X-1. \cite{2012PASJ...64...72S} obtained the following $N_H$ estimate from the best fit to the Aql~X-1 soft-state X-ray spectrum in outburst observed by \suzaku:
\begin{equation}
N_H = (3.6\pm0.01)\times10^{21} ~\frac{atoms}{cm^{2}} ~.
\label{eq:Nh}
\end{equation}  
Note, that this value well agrees with hydrogen column density within Galaxy in the direction of Aql~X-1 estimated from \citep{1990ARAA..28..215D}.

Then, we estimated the color excess $E_{B-V}$ coefficient in the direction to Aql~X-1 by using the recalibrated Galaxy extinction maps based on dust emission measured by COBE/DIRBE and IRAS/ISSA \citep{2011ApJ...737..103S}: $E_{B-V}\approx0.65^{mag}$. On the other hand $E_{B-V}$ can be estimated using common $N_H$--$A_V$ relation between $V$-band optical extinction and the hydrogen column density. Using for $N_H/A_V$ a classical estimate from \cite{1995AA...293..889P} for the usual extinction law with parameter $R_V=A_V/E_{B-V}=3.1$, from (\ref{eq:Nh}) one can obtain the following estimate for Aql~X-1 color excess:
\begin{equation}
E_{B-V}\approx0.65^{mag} ~,
\label{eq:Ebv_best}
\end{equation}
which coincides with both Galaxy extinction derived from dust emission map \citep{2011ApJ...737..103S}. It is worth noting, that in \cite{1999AA...347L..51C} the close value of color excess for Aql~X-1 has been measured from optical spectroscopy of the optical counterpart during the SXT quiescence state: $E_{B-V}=0.5\pm0.1^{mag}$ for both Aql~X-1 optical counterpart and its close interloper star (authors assume that both stars are reddened by the same amount) from joint spectral models fitting. Note, that \cite{1978ApJ...220L.131T} mentions that a nearby (1'.4) B5 V star lies at a distance 10\,kpc, well above the Galactic dust layer, has an optical reddening $E_{B-V}\approx0.73^{mag}$. 

In this work we adopt (\ref{eq:Nh}) and (\ref{eq:Ebv_best}), as a best estimates for interstellar extinction in the direction to Aql~X-1.

\begin{table*}
\begin{center}
\caption[\swiftxrt observations of Aql X-1]{\swiftxrt observations of Aql X-1 during outburst rise in 2013.} \label{tbl:xrt_all} 
\begin{tabular}{cccccc}
\hline\hline
Obs Id        &  Tstart,   & Exposure, & $F_{\rm X,0.5-10}$,                     & $kT$,  & Photon  \\
                  & MJD        &  ks      &  $10^{-9}$ erg~s$^{-1}$~cm$^{-2}$ & keV  & index  \\
\hline
Hard state \\
00035323003    & 56451.5371  &  0.85    &  $ 0.343\pm 0.021$    & $0.6^{+0.4}_{-0.1}$ &  $1.3^{+0.2}_{-0.3}$ \\
00035323004    & 56452.4042  &  0.80    &  $ 0.506\pm 0.010$    & $0.72\pm0.05$    &  $1.48\pm0.06$ \\
00035323005    & 56453.6179  &  1.00    &  $ 0.931\pm 0.013$    & $0.82\pm0.05$    &  $1.44\pm0.04$ \\
00035323006    & 56454.6236  &  0.95    &  $ 1.552\pm 0.016$    & $1.07\pm0.05$    &  $1.59\pm0.04$ \\
00035323007    & 56456.8132  &  1.49    &  $ 3.565\pm 0.025$    & $1.21\pm0.06$    &  $1.53\pm0.03$ \\
00035323009\_1 & 56457.0971  &  0.37    &  $ 3.776\pm 0.043$    & $1.62\pm0.07$    &  $1.86\pm0.07$ \\
00035323009\_2 & 56457.6291  &  0.54    &  $ 4.875\pm 0.045$    & $1.65\pm0.09$    &  $1.69\pm0.04$ \\
00035323009\_3 & 56457.9627  &  0.80    &  $ 5.916\pm 0.041$    & $1.50\pm0.06$    &  $1.61\pm0.03$ \\
\hline
Soft state \\
00035323008\_1 & 56458.8272  & 1.07   &  $17.100\pm 0.100$    & $0.80\pm0.01$    &  $1.66\pm0.02$ \\
00035323008\_2 & 56458.9693  & 0.36   &  $17.660\pm 0.144$    & $0.76\pm0.02$    &  $1.60\pm0.03$ \\
00035323010    & 56459.5475  & 1.54   &  $25.293\pm 0.092$    & $0.92\pm0.01$    &  $1.54\pm0.02$ \\
00035323011    & 56460.2154  & 1.48   &  $27.102\pm 0.125$    & $1.06\pm0.01$    &  $1.67\pm0.02$ \\
\hline
\end{tabular}
\end{center}
\vspace{3mm}
\begin{tabular}{ll}
\end{tabular}
\end{table*}

\begin{table}
  \begin{center}
    \caption[\swiftuvot observations of Aql X-1]{\swiftuvot observations of Aql X-1 during outburst rise in 2013} \label{tbl:uvot_all} 
    \begin{tabular}{cccc}
      \hline\hline
      ObsID/Filter &  Tstart &  Exposure & Magnitude  \\
                   &   MJD   &  s        &          \\
      \hline
      \multicolumn{1}{c}{00035323003} \\
                   B    & 56451.5425 &  211.4 &  $18.60\pm0.12$ \\
                   U    & 56451.5399 &  211.4 &  $19.47\pm0.31$ \\
                   UVW1 & 56451.5349 &  423.0 &  $19.53\pm0.26$ \\
                   UVW2 & 56451.5451 &  116.3 &  $19.63\pm0.46$ \\
      \hline
      00035323004 \\ 
                    B    & 56452.4109 &  155.5 &  $18.64\pm0.14$ \\
                    U    & 56452.4084 &  205.5 &  $18.66\pm0.17$ \\
                    UVW1 & 56452.4036 &  411.2 &  $19.27\pm0.21$ \\
      \hline
      00035323005 \\
                    B    & 56453.6216 &  118.9 &  $18.55\pm0.15$ \\
                    U    & 56453.6201 &  118.9 &  $18.54\pm0.20$ \\
                    V    & 56453.6287 &    4.0 &  $17.22\pm0.72$ \\
                    UVW1 & 56453.6173 &  238.9 &  $19.58\pm0.35$ \\
                    UVW2 & 56453.6231 & 478.1  &  $20.52\pm0.45$ \\
      \hline
      00035323006 \\
                    B    & 56454.6258 &   77.5 &  $18.31\pm0.16$ \\
                    UVM2 & 56454.6314 &  222.6 &  $20.09\pm0.56$ \\
                    U    & 56454.6248 &   77.5 &  $18.58\pm0.25$ \\
                    V    & 56454.6305 &   77.5 &  $17.25\pm0.16$ \\
                    UVW1 & 56454.6230 &  154.3 &  $19.36\pm0.37$ \\
      \hline 
      00035323007 \\
                    U    & 56456.8125 & 1469.9 &  $17.76\pm0.04$ \\
      \hline
      00035323009 \\
                    UVW2 & 56457.0964 &  362.4 &  $20.38\pm0.48$ \\
                    UVW2 & 56457.6286 &  521.1 &  $19.05\pm0.15$  \\
                    UVW2 & 56457.9620 &  810.1 &  $19.13\pm0.12$ \\
      \hline
      00035323008 \\
                    UVM2 & 56458.8265 & 1051.7 &  $19.33\pm0.15$  \\
                    UVM2 & 56458.9686 &  365.0 &  $19.80\pm0.36$  \\
      \hline
      00035323010 \\
                    UVW1 & 56459.5468 & 1513.4 &  $17.68\pm0.04$  \\
      \hline
      00035323011 \\
                    U    & 56460.2147 & 1463.7 &  $16.48\pm0.03$  \\
      \hline 
      \hline
    \end{tabular}
  \end{center}
\label{tbl:uvot_log}
\end{table}

\section{Observations and data reduction}
\label{sec:reduction}
 
\subsection{MAXI} 
We downloaded daily- and orbit-averaged light curves of Aql X-1 from MAXI
Archive official website\footnote{see http://maxi.riken.jp/top/}. For
counts-to-flux conversion the Crab spectrum was assumed in the efficiency
correction for each band. Fluxes in Crab units for MAXI instrument were obtained using standard
conversions: 1 Crab approximately equals to 3.6 ph/s/cm$^2$ in the total
$2-20$\,keV band and 1.87, 1.24, 0.40 ph/s/cm2 for 2-4, 4-10, 10-20 keV band, respectively. 
In order to obtain more accurate luminosities from instrumental count rates in the $2-10$~keV band, we derived appropriate
conversion factor by using an overlapping series of XRT/Swift $2-10$~keV flux
measurements in the time interval 56456-56461~MJD ($\pm$3 days around state
transition during outburst rise). The derived count rate--flux conversion factor appeared to be
close (only $+15$\% correction) to standard conversion
($1~Crab(2-10~keV)=3.11~counts~cm^{-2}~sec^{-1}=2.156\cdot10^{-8}~erg~s^{-1}~cm^{-2}$).

\subsection{\swift} 

\swift observatory \citep{2004ApJ...611.1005G} provides
possibility to get the simultaneous broadband view from optical to hard
X-rays, that is crucial for X-ray Novae studies. In this work we used
observations covering the rising phase of Aql X-1 outburst --- between 56450 and
56462 MJD (altogether 12 snapshot observations). Tables \ref{tbl:xrt_all} and \ref{tbl:uvot_all} provides a journal of observations carried out by XRT and UVOT instruments, respectively.  Below we review Swift data reduction in detail.
 
\subsubsection{\swiftxrt}
XRT observed Aql X-1 both in Windowed Timing (WT) mode, while the transient was bright, and Photon Counting (PC) mode, for low count rate snapshots. The data were processed using tools and packages available in {\tt FTOOLS/HEASOFT 6.14}.  Initial cleaning of events has been done using {\tt{}xrtpipeline} with standard parameters. The following analysis was performed as described in \cite{2009MNRAS.397.1177E}. In particular, for the PC mode data, radius of the circular aperture for the source extraction was depending on the count rate ranging from 5 to 30 pixels \citep{2009MNRAS.397.1177E}; for the WT mode data, radius of the source extraction region was 25 pixels. Background was extracted from the annulus region with the inner (outer) radius of 60 (110) pixels in both PC/WT observational modes. In the case of pile up, central region of the source was excluded to ensure the final count rate below 0.5 and 100 counts s$^{-1}$ for the PC and WT modes, correspondingly. The obtained spectra were grouped to have at least 20 counts bin$^{-1}$ using the FTOOLS {\tt grppha}. To avoid any problems caused by the calibration uncertainties at low energies\footnote{http://www.swift.ac.uk/analysis/xrt/digest\_cal.php}, we restricted our spectral analysis to the $0.5-10$ keV. 
   
In this work we used \swiftxrt observations obtained during the outburst rise phase only (9 pointing observations containing 12 snapshots). The standard spectral analysis of the XRT data was performed. We successfully fitted (with $\chi^2_r\approx1$) object X-ray spectrum in each snapshot by phenomenological $phabs*(diskbb+powerlaw)$ model in XSPEC package. The interstellar absorption parameter was fixed to the standard Aql~X-1 value (see Table~\ref{tbl:aqlx1_pars}). Finally, we derived $0.5-10$~keV fluxes for all available {\it Swift}/XRT snapshot observations,  and present them in the Table~\ref{tbl:xrt_all}, together with best-fit parameters for adopted spectral models. Errors reported in the Table~\ref{tbl:xrt_all} are purely statistical and correspond to 1$\sigma$ confidence level. However, ARF calibration uncertainties for the {\it Swift}/XRT instrument can reach 10\%\footnote{Swift Helpdesk private communication} but wasn't included into our analysis.

\subsubsection{\swiftbat}
\swiftbat detector provided a hard X-ray measurements of the outburst
light curve.  We downloaded daily- and orbit-averaged light curves of Aql X-1
from \swiftbat Hard X-ray Transient Monitor archive website
\footnote{http://swift.gsfc.nasa.gov/results/transients/index.html}.
For counts-to-flux conversion, it was assumed that 1 Crab equals to 0.220
counts/s/cm$^2$ in the $15-50$ band. The $15-50$~keV BAT fluxes and were derived from BAT
count rate using standard conversion:
$1~Crab(15-50~keV)=0.22\,ph/s/cm^2=1.345\cdot10^{-8}~erg~s^{-1}~cm^{-2}$.

\subsubsection{\swiftuvot}
The \swiftuvot observation log is shown in the Table~\ref{tbl:uvot_all}. UVOT exposures were taken in six filters (V, B, U, UVW1,UVW2, and UVM2)
for the first four observations and with the ``filter-of-the-day'' subsequently. Errors reported in the Table~\ref{tbl:uvot_all} are purely statistical and correspond to 1$\sigma$ confidence level.

For the data reduction images initially preprocessed at the Swift Data Center at the Goddard Space Flight Center were used. Subsequent analysis has been done following procedure described at the web-page of UK Swift Science Data Centre.\footnote{http://www.swift.ac.uk/analysis/uvot/index.php} Namely, photometry was performed with {\tt uvotsource} procedure with source apertures of radius 5 arc seconds and 10 arc seconds for the background for all filters.  Finally, spectral files for fitting in {\tt XSPEC} were produced with the {\tt uvot2pha} procedure.

It can be noted, that a 5" aperture contains flux from the group of faint stars, located nearby to Aql X-1 counterpart.  By applying the background subtraction procedure we were able to eliminate the contamination from nearby stars and Aql~X-1 quiescent light (see \S\ref{sec:data-optical} for detail).

\begin{figure}    
  \includegraphics[width=\columnwidth]{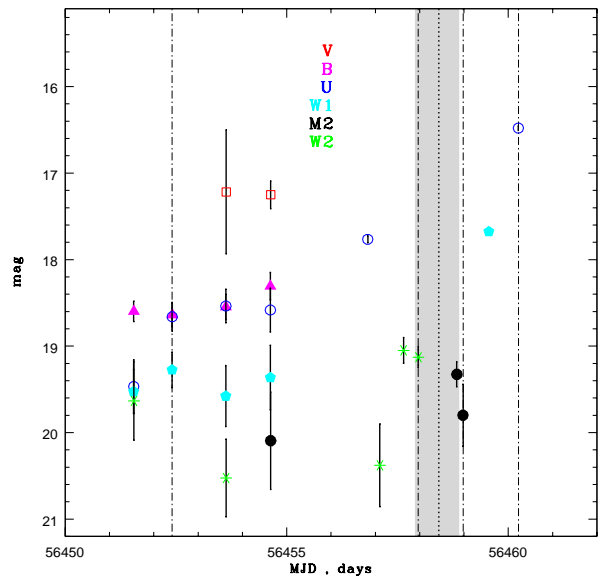}
  \caption{\swiftuvot light curve during outburst rise in Aql~X-1. The grey shaded band marks the time interval of hard/soft X-ray state transition. The state transition midpoint is shown by doted vertical line. Dot-dashed vertical lines correspond to time moments of broad-band SED measurements (see \S\ref{sec:broadband-SED-points}).}
    \label{fig:lcurve_uvot}
\end{figure}    

\begin{figure*}    
  \includegraphics[width=17cm]{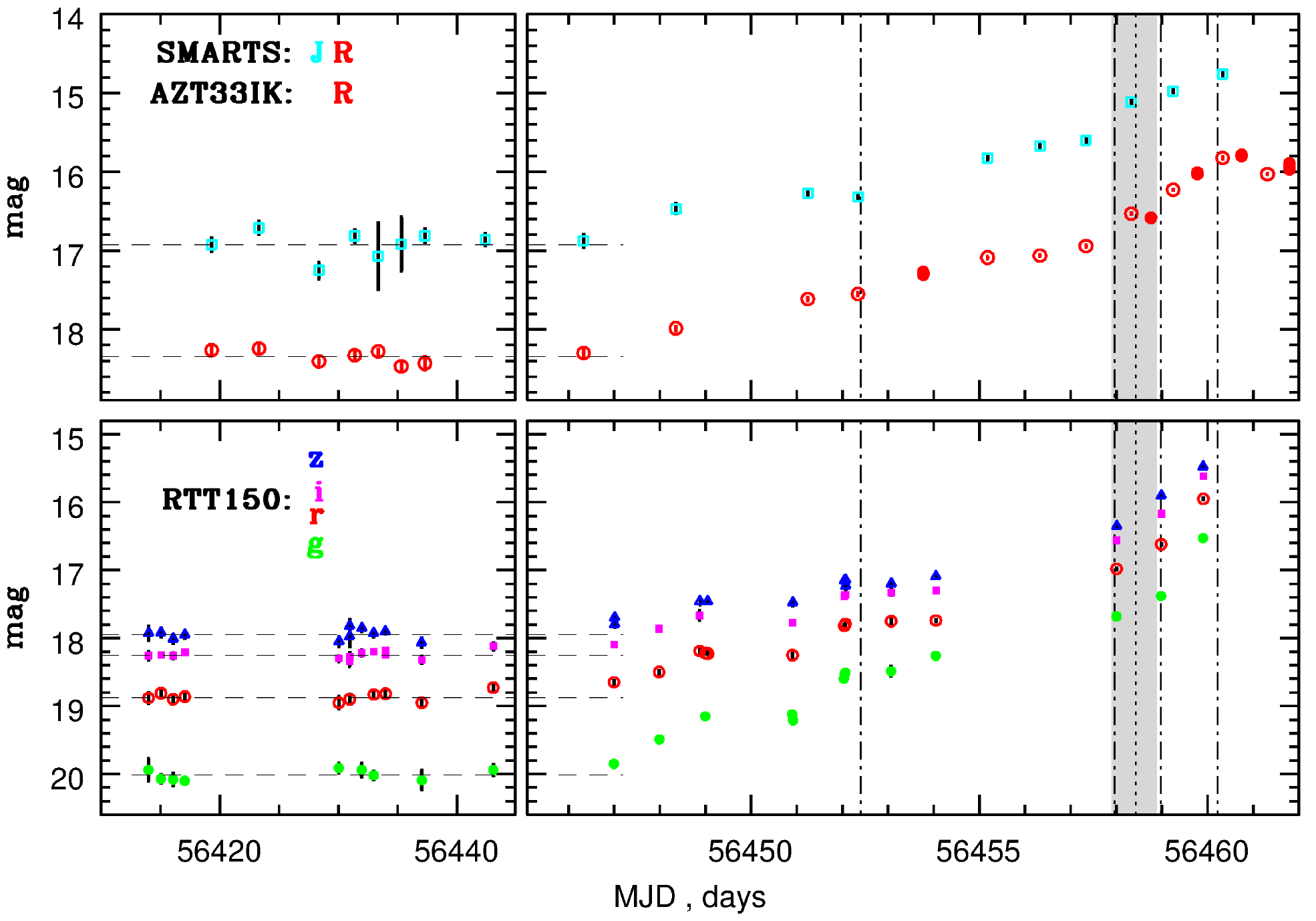} 
 \caption{Optical-NIR light curves during the Aql~X-1 quiescence (left panels) and rise phase (right panels) of major outburst in 2013 from RTT150, AZT33IK and SMARTS telescopes. The derived Aql~X-1 quiescent fluxes in $g$, $r$, $i$, $z$, $R$ and $J$ bands are shown by horizontal dashed lines. The grey shaded band mark the time interval of hard/soft X-ray state transition. Dot-dashed vertical lines correspond to time moments of UVOT snapshot observations.}
  \label{fig:lcurve_opt} 
\end{figure*}

\begin{table}
  \begin{center}
    \begin{tabular}{cccccc}
      \hline \hline 
      filter & $\lambda_{eff}$ & FWHM &  $F_{\nu,0}$ & Ref & Aql~X-1$^\star$\\
             & \AA                     & \AA       &  Jy                &      & in quiescence \\
      \hline    
      \multicolumn{6}{l}{\swiftuvot} \\ 
      \hline         
      W2 & 1928  &  657 &  738   &  [1]  & $24.16\pm0.86$\\
      M2 &  2246 &  497 &  766   &  [1]  & $23.0$$^{\star\star}$ \\
      W1 &  2600 &  693 &  904   &  [1]  & $22.47\pm0.26$ \\
       U   &  3465 &  785 & 1419  &  [1]  & $21.50\pm0.25$ \\ 
       B   &  4392 &  975 & 4093  &  [1]  &  $20.03\pm0.20$ \\ 
       V   &  5468 &  769 & 3631  &  [1] &  $18.39\pm0.12$ \\
      \hline       
      \multicolumn{6}{l}{\rtt, \smarts}  \\
      \hline       
      g'   &  4714 & 1379 & 3631  & [2] & $20.02\pm0.07$ \\
      r'    &  6182 & 1382 & 3631  & [2] & $18.88\pm0.05$ \\ 
      i'    &  7592 & 1535 & 3631  & [2] & $18.26\pm0.05$ \\
      z'   &  9003 & 1370 & 3631  & [2] & $17.90\pm0.18$ \\    
       R &  6410 & 1576 & 3064  & [3,4] & $18.35\pm0.08$ \\
       J & 12600 & 2000 & 1603  & [5] & $16.93\pm0.17$  \\
      \hline \hline 
    \end{tabular}
  \end{center}
[1] \cite{2008MNRAS.383..627P} , [2] \cite{1996AJ....111.1748F} ,
  [3] \cite{1998ApJ...500..525S}, [4] \cite{1998AA...333..231B}, [5] \cite{1985AJ.....90..896C} \\
 $^\star$Aql~X-1 quiescent flux with interloper star (see text):
{\it UVOT ---} sum of all available data (photometry within $5^{\prime\prime}$ aperture) for time intervals 15.03-15.11.2012 and 15.09-15.11.2013. {\it RTT150,SMARTS ---} sum of all available data (PSF photometry, see \S\ref{sec:data-optical}) for pre-outburst period 27.04-27.05.2013. \\
$^{\star\star}$ Aql~X-1 quiescent flux in UVOT M2 band was derived from interpolation between W2 and W1 values.\\
\caption{Average flux levels for Aql~X-1 in quiescence (with interloper star) were measured at the period of XN quiescence in 2012 (UVOT data) and the pre-outburst period in 2013 (\rtt and \smarts data).}
  \label{tbl:ph_sys}
\end{table}

\subsection{Ground-based optical data}
\label{sec:data-optical}
The Aql~X-1 optical counterpart lies in a crowded field with 4 nearby interloper stars separated from Aql~X-1 star only by 0''.48, 2''.6, 2''.4 and 1''.3, respectively \citep{1999AA...347L..51C,2012ApJ...749....3H}, which may produce contamination. The 0''.48 interloper star  is substantially brighter ($V=19.4^{mag}$) than Aql~X-1 optical counterpart in the quiescence state ($V=21.6^{mag}$). Once an outburst begins, photons from Aql~X-1 became dominant. In order to obtain a correct flux for Aql~X-1 counterpart in outburst, we subtracted an average flux levels measured during the period of X-ray Nova quiescence in 2012 and the pre-outburst period in 2013 (see Table~\ref{tbl:ph_sys}). The optical data reduction procedure is described below.

In Apr-Nov 2013 the following small-size ground-based optical telescopes
have participated in the multi-wavelength monitoring campaign of Aql X-1:
\begin{itemize} 
\item \rtt --- the joint Russian-Turkish 1.5-m Telescope (30$^\circ$19'59.9''E, 36$^\circ$49'31.0''N, 2538.6-m above sea-level, T{\"U}BITAK National Observatory, Turkey) equipped with the TFOSC focal instrument for direct imaging and spectral observations.  The object was observed in $g'$, $r'$, $i'$, $z'$ bands. 
\item \azt 1.6-m telescope (100$^\circ$55'13'' E, 51$^\circ$37'18.10'' N, 2000-m above sea-level, Sayan Observatory, Russia). For direct
  imaging and fast photometry, a sCMOS Andor camera was used. The object was observed in $R$ band.
\item \zeiss 1-m telescope (41$^\circ$26'30'' E, +43$^\circ$39'12'' N, 2070-m above sea-level, Special Astrophysical Observatory, Russia). The object was observed in $R$ band, monitoring observations started after the outburst maximum in 2013 (data from this telescope we will not discuss in this work).
\item \smarts 1.3-m telescope at Cerro Tololo (Chile). Aql X-1 was monitored in $R$ and $J$-bands at the regular basis. We used a publicly available\footnote{www.astro.yale.edu/smarts/xrb/home.php} \smarts light curves in our analysis. The photometric reduction procedure were performed by Yale SMARTS XRB team, following closely the reduction steps described in \cite{2012AJ....143..130B}.
  \end{itemize}

As it was emphasised in previous optical variability studies of Aql~X-1 (see e.g. \cite{2000AJ....120..943W}), the use of point-spread functions to extract the source counts (instead of ordinary aperture photometry) is crucial to obtain reliable optical flux measurements for Aql~X-1 optical counterpart. For the photometric observations carried out at \rtt, \azt, \zeiss telescopes, we extracted instrumental magnitudes for Aql~X-1 and few local comparison stars (see below) using the DAOPHOT routine \citep{1987PASP...99..191S} in the Interactive Data Language (IDL). We used two iterations of the point-spread function fitting routine; a third iteration did not improve the precision of the photometry. Photometric fluxes of Aql~X-1 in standard R,g',r',i',z' bands (see Table~\ref{tbl:ph_sys}) were obtained from instrumental counts by using the following secondary standards located nearby in the Aql~X-1 field:  (i) $\alpha_1= 287.8073766^\circ$, $\delta_1=0.5811534^\circ$;  (ii) $\alpha_2=287.8032941^\circ$, $\delta_2=0.5781298^\circ$; (iii) $\alpha_3=287.8179977^\circ$, $\delta_3=0.5759676^\circ$;  (iv) $\alpha_4=287.8082571^\circ$, $\delta_4=0.5778305^\circ$; (v) $\alpha_5=287.8204860^\circ$, $\delta_2=0.5873818^\circ$. These local comparison stars are invariable (within statistical uncertainties) during the whole period of Aql~X-1 monitoring observations and have visual $R$ magnitudes in the range $15^{mag}\div17^{mag}$. Their R,g',r',i',z' fluxes in the standard (see Table~\ref{tbl:ph_sys}) photometric system  were derived by observation of the Aql~X-1 field and primary standard stars \citep{2002AJ....123.2121S,1992AJ....104..340L} during observation in Nov 2013 in one of nights with photometric atmospheric conditions. We conservatively estimated the final accuracy of absolute photometric calibration for \rtt, \azt, \zeiss and \smarts telescopes as 3\%. 

In this paper \rtt g'r'i'z' flux measurements for Aql~X-1 are presented in AB photometric system, all other (UVOT, R, J) flux measurements are presented in Vega system. The adopted effective wavelength, bandwidth and photometric zero points for all used filters/instruments are shown in the Table~\ref{tbl:ph_sys}.
 
\section{Outburst rise in Aql~X-1}
\label{sec:results}
New outburst in the Aql~X-1 X-ray Nova system was detected 3 June 2013 \citep{2013ATel.5114....1M} during the campaign of optical monitoring observations of the object, started in April~2013 at 1.5-m Russian-Turkish telescope \rtt.  At Figures~\ref{fig:lcurve_uvot} and \ref{fig:lcurve_opt} (right panels) all available NUV, Optical and NIR light curves obtained during the rising phase of Aql~X-1 outburst at \swiftuvot, \rtt, \azt and \smarts instruments are shown. At Figure~\ref{fig:lcurve_opt} (left panels) we present the available pre-outburst Optical-NIR light curves from ground-based telescopes. The horizontal dashed lines mark the measured background level (which is dominated by the close interloper star (see \S\ref{sec:data-optical}). 

\begin{figure}    
  \includegraphics[width=\columnwidth]{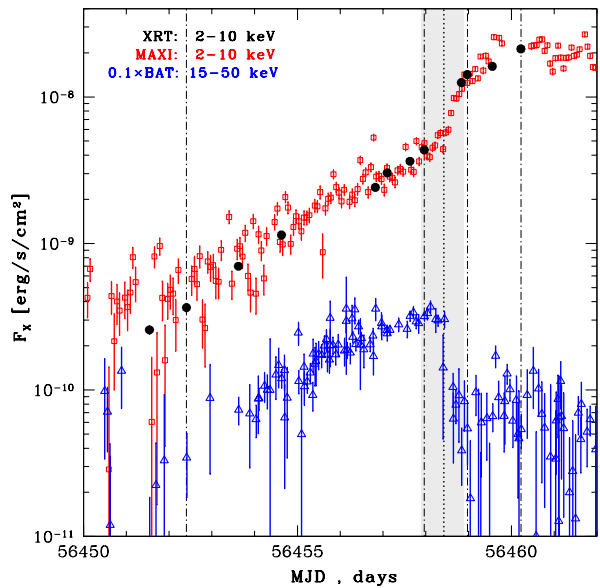}
  \caption{Orbit-averaged \maxi ($2-10$\,keV) and \swiftbat ($15-50$\,keV) light curves are shown together with
    \swiftxrt light curve (in the $2-10$\,keV band). The midpoint and duration of hard/soft X-ray state transition are shown by doted vertical line and grey shaded band. Dot-dashed vertical lines correspond to time moments of broad-band SED measurements.}
    \label{fig:lcurve_Xray}
\end{figure}

In order to measure the broadband NUV-NIR spectral evolution during outburst rise period in Aql~X-1, we chose four characteristic time moments, where observations from two instruments (\swiftuvot--\rtt or \swiftuvot--\smarts) were carried out quasi-simultaneously (within the time interval $\Delta{}t\lesssim0.1^d$). These time moments are marked at Figures~\ref{fig:lcurve_uvot},\ref{fig:lcurve_opt},\ref{fig:lcurve_Xray} by vertical dot-dashed lines (the corresponding broad-band NUV-NIR SEDs will be discussed in the section \S\ref{sec:broadband-SED-points} below).

\begin{figure} 
  \includegraphics[width=\columnwidth]{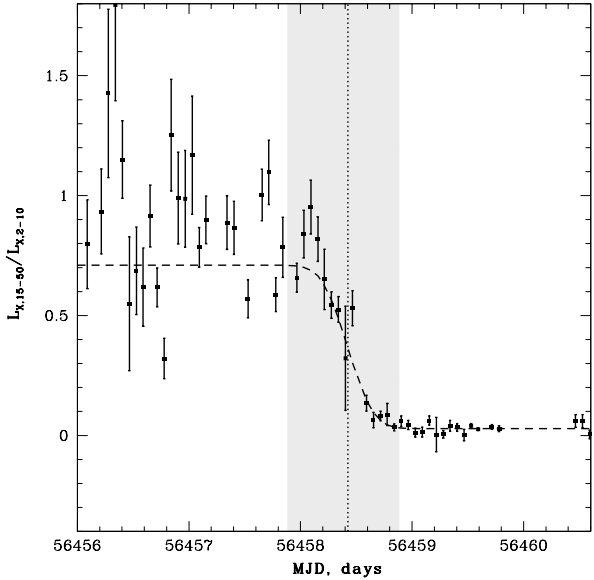}
  \caption{X-ray color ($15-50$/$2-10$\,keV) evolution around the
    hard/soft X-ray state transition interval during the rising part of
    Aql~X-1 outburst. The grey shaded band shows the estimated time interval
    of state transition, the midpoint is shown by doted vertical line. The
    best fitted model to X-ray color evolution during state transition is
    shown by dashed line.}
  \label{fig:hs_transition}    
\end{figure}    

After the outburst detection in optical g', r', i', z'-bands, the accretion activity of the source was soon confirmed by \swiftxrt follow-up observations \citep{2013ATel.5117....1D}. The X-ray outburst happened to be among the brightest in soft X-rays among all Aql~X-1 accretion events observed by \maxi or \rxteasm All Sky Monitors since 1997 \citep{2014MNRAS.439.2717G}.  The overall morphology of this outburst in soft X-rays is characterized by a fast ($\sim10^d$) rise and a long ($\sim50^d$) decay. This type of curves are often observed in X-ray Novae \citep{1997ApJ...491..312C} and called FRED (Fast-Rise-Exponential-Decay). The FRED-type light curves in SXT are well qualitatively reproduced by standard Disc Instability Model (DIM), if effects of accretion disk evaporation and irradiation by the central source are taken into account \citep{2001AA...373..251D}. The orbit-averaged light curves from \maxi ($2-10$~keV) and \swiftbat ($15-50$~keV) for the Aql~X-1 outburst rise phase are shown at the Figure~\ref{fig:lcurve_Xray}. At the same Figure, we show all available X-ray pointing measurements (in the same soft energy range $2-10$\,keV), carried out by \swiftxrt telescope during this interval. 
 
The remarkable drop ($\sim5$ times decrease at a time scale $<1^d$) of hard X-ray flux, when the soft X-ray brightness is still rising, corresponds to the time moment of hard/soft X-ray state transition.  At Figure~\ref{fig:hs_transition} the (15-50\,keV)/(2-10\,keV) X-ray color
evolution during the interval of state transition is shown. From this data, one can measure the midpoint and duration of
state transition, by fitting the X-ray color evolution by the appropriate low-parametric model. We chose the following function
\begin{equation}
c(t)= p_0 + p_1\times\bigg[erf\bigg(\frac{t-p_2}{p_3}\bigg)-1\bigg] ~,
\label{eq:c_model}
\end{equation} 
where $erf(x)$ is the error function in its standard form $erf(x)=\frac{2}{\sqrt{\pi}}\int_0^x e^{-\xi^2}d\xi$, and $p_0$, $p_1$,
$p_2$, $p_3$ are free parameters. The state transition itself we defined as a time interval, where the majority (99.7\%) of color change take place (according to our best fit, see dashed line at Figure~\ref{fig:hs_transition}). We obtained the state transition midpoint and duration for the Aql~X-1 outburst rise in 2013:
\begin{eqnarray}   
  T_{h/s} &=& p_2 = 56458.425 ~[MJD] ~,\\ \nonumber
  \Delta{}T_{h/s} &=&\frac{4}{\sqrt{2}}\times{}p_3 = 1.073~[day]
\label{eq:tx}
\end{eqnarray}
and show them at Figures~\ref{fig:lcurve_uvot}--\ref{fig:hs_transition} by doted vertical line and grey shaded band, respectively.  It is worth noted, that the interval of fast changes in hard and soft X-ray fluxes during state transition is even smaller than the value $\Delta{}T_{h/s}$ defined above. One can estimate from Figure~\ref{fig:lcurve_Xray}, that fast increase (decrease) of soft(hard) X-ray flux begins about the transition midpoint $T_{h/s}$. Thus we can estimate the actual duration of fast changes in hard (soft) X-ray fluxes during state transition as $\approx\Delta{}T_{h/s}/2$, respectively.
We defined a convenient time variable $t$, measured with respect to the state transition midpoint:
\begin{equation} 
t = T - T_{h/s} ~.
\end{equation}

In our Swift/XRT observations we are able to measure accurately only the soft fraction $F_{\rm X,0.5-10}$ of the total X-ray flux $F_{\rm X,bol}$ (which we define here in the energy range $0.5-100$\,keV). The bolometric X-ray flux from the inner parts of the disc can be estimated as:
\begin{equation}
\label{eq:Fxbol}
F_{\rm X,bol} = f_{bol} \cdot F_{\rm X,0.5-10} ~,
\end{equation}
where $f_{bol}$ means a bolometric correction coefficient. Note, that the bolometric correction is substantial for the spectrum in the hard X-ray state. To estimate $f_{bol}$ we used results from \cite{2012PASJ...64...72S}, who analyzed broad-band X-ray observations in the hard and soft X-ray states during Aql~X-1 outburst in Sep-Oct~2008, carried out by \suzaku observatory. By using their best-fit models in Table~2 and 3 (with fixed $N_H=0.36\cdot10^{22}$cm$^{-2}$), we calculated $0.5-10$\,keV, $2-10$\,keV, $15-50$\,keV and "bolometric" $0.5-100$\,keV unabsorbed fluxes for the typical soft and hard X-ray state spectra. The estimated bolometric corrections are $f^{hard}_{bol}=1.96$ and $f^{soft}_{bol}=1.08$ for observational data points before and after X-ry state transition, respectively. In addition we derived the $F_{X,15-50}/F_{X,2-10}$ ratio: $0.89$ and $0.036$ in the hard and soft state, respectively. As can be seen at Figure~\ref{fig:hs_transition}, these values are well compared to the observed BAT/MAXI X-ray colors before and after state transition. 

\subsection{Broad-band SED measurements}
\label{sec:broadband-SED-points}

There are two time moments before X-ray state transition midpoint and  two time moments after, when we are able to measure a quasi-simultaneous (within $\lesssim0.1^d$) broad-band SED of the source. Below we describe derived SED measurements and the fitting procedure in detail. Broad-band SED measurements during the outburst rise in Aql~X-1:
\begin{enumerate}
\item $t\approx-6.02^d$. At this time moment, \swiftxrtuvot observations were carried out quasi-simultaneously with \smarts telescope ($t=-6.08^d$), and we combined these data to construct broad-band SED. Additionally, as can be noted (see Figures~\ref{fig:lcurve_uvot}--\ref{fig:lcurve_opt}), the subsequent \swiftuvot observation at $t=-4.80^d$ shows the same (within uncertainties) NUV fluxes. Thus we included this \swiftuvot observation and \rtt observation carried out in between at $t=-5.35^d$ into the combined SED. The derived SED is shown at the Figure~\ref{fig:sed:124} (left panel), where all the "non-simultaneous" data points from \rtt and the second \swiftuvot observation are shown by open symbols. \label{sed:1}
\item $t\approx-0.46^d$. This time moment immediately before state transition, when \swiftuvot W2-band observation at $t=-0.46^d$ was carried out quasi-simultaneously with \rtt ($t=-0.42^d$). As the previous \swiftuvot observation at $t=-0.79^d$ shows the same (within uncertainties) W2 flux, we decided to include it into the combined SED. The resulting SED is shown at Figure~\ref{fig:sed:124} (central panel), the "non-simultaneous" \swiftuvot data point is shown by open symbol.  \label{sed:2}
\item $t\approx+0.55^d$. This is the most interesting SED measurement, we luckily obtained it immediately after hard/soft X-ray state transition, the time moment  of \swiftuvot M2-band observation was carried out quasi-simultaneously with \rtt ($t=+0.56^d$). As the previous \swiftuvot observation at $t=+0.41^d$ shows the same (within uncertainties) M2 flux, we decided to include it into the combined SED. The resulting SED is shown at Figure~\ref{fig:sed:124} (central panel), where the "non-simultaneous" \swiftuvot data point is shown by open symbol. \label{sed:3}
\item $t\approx+1.80^d$. This  is the final SED measurement we obtained near the outburst maximum in X-rays (see Figure~\ref{fig:lcurve_Xray}).  The \swiftuvot U-band observation was carried out quasi-simultaneously with \smarts ($t=+1.91^d$). The resulting SED is shown at Figure~\ref{fig:sed:124} (right panel). \label{sed:4}
\end{enumerate}

The SED fitting procedure was performed in XSPEC package \citep{1996ASPC..101...17A}, which provides a framework to compare various theoretical spectral models with observed spectra (primary in the X-ray domain). XSPEC can be successfully used to fit spectral data from IR/Optical/UV observations \citep{2010HEAD...11.0905A}. We converted all NUV, Optical and NIR photometric measurements into pha-files using procedure flx2xsp from FTOOLS package. For all filters, responses were defined by flat transmission curves with parameters $\lambda_{eff}$ and $\Delta\lambda$ (FWHM) (see Table~\ref{tbl:ph_sys}). We note that the observed fluxes contain Aql~X-1 counterpart and nearby $0.48^{\prime\prime}$ interloper star for ground-based Optical-NIR observations, and all nearby stars within  $5^{\prime\prime}$ aperture for \swiftuvot observations. In order to investigate the spectral evolution of Aql~X-1 counterpart in outburst, we subtracted the corresponding flux levels measured during the period of Aql~X-1 quiescence (see Table~\ref{tbl:ph_sys}).

The interstellar extinction in photometric filters was calculated by using REDDEN model in XSPEC. This model utilize \cite{1989ApJ...345..245C} extinction law from far-IR to far-UV as a function of wavelength and of the parameter $E_{B-V}$. For all spectral fits below we adopted the fixed color excess value $E_{B-V}=0.65^{mag}$, as a best estimate for Aql~X-1 (see \S\ref{sec:aqlx1}). 
 
\subsection{Adopted spectral models}
\label{sec:broadband-SED-models}
We tried to fit Aql~X-1 NUV-NIR SEDs by two low-parametric spectral models:
\begin{enumerate}[label=(\Alph*)]
\item {\it Absorbed blackbody emission} ($redden*bbodyrad$),  \label{m:bb}
\item {\it Absorbed emission from multi-color disc} with possible X-ray irradiation ($redden*diskir$).  \label{m:disc}
\end{enumerate}
Our choice of spectral models (A) and (B) is physically motivated.  The simplified analytical picture of the non-stationary disc accretion during outburst rise phase in X-ray Binaries was proposed in the work of \cite{1987SvAL...13..386L}. The accretion disc development from the initial ring of matter can be divided into 3 characteristic stages:
\begin{enumerate}[label=\Roman*]
\item Formation of the disc from the initial ring of matter ("torque" formation stage).
\item Quasi-stationary accretion with increasing accretion rate. At this stage a radially constant accretion rate is established in the inner regions of accretion disc. Near outer radii of the disc no changes from the initial mass distribution are expected and a transition zone is developed at intermediate radii. The region of quasi-stationary solution continuously expands as the transition zone moves outward. \label{stage:2}
\item The accretion attenuation phase after the outburst maximum. 
\end{enumerate}
We are interested in stage \ref{stage:2}, which could be potentially observed by our broad-band observations of outburst rise in Aql~X-1 system. During this stage the mass distribution in the outer regions of the disc transforms from initial distribution (at the pre-outburst quiescence) into the stationary accretion disc (near the outburst maximum). Accordingly, the spectral evolution in the NUV-NIR range (which corresponds to emission from the outer parts of the disc) should transform from a single-temperature blackbody emitting ring into the multi-colour (irradiated or non-irradiated) accretion disc emission. The initial ring of matter in the  \citep{1987SvAL...13..386L} analytical model can be in reality a manifestation of the accretion disc with a surface density profile, highly concentrated to some outer radius --- like $\Sigma\propto{}R^{1.14}$, which is supposed to form in the disc during the X-ray Nova quiescence (see \cite{2001NewAR..45..449L}). The present numerical models of XN outbursts also show, that the single temperature emission remains at early stages of SXT outburst (see e.g. Figure~5 in \cite{2001AA...373..251D}). Note that, alternatively, a single blackbody model may correspond to the emission from the X-ray heated surface of companion star (if X-ray irradiation is strong enough) or a hot point, where a stream from L1-point meet the accretion disc. At the end of stage \ref{stage:2}, the multi-color disc model corresponds to emission from the standard \citep{1973AA....24..337S} steady-state optically-thick accretion disc with possible X-ray irradiation. The multi-color disc emission is expected to be established about the moment of the outburst maximum, the radial mass distribution in the disc at that moment does not depended on initial mass distribution in pre-outburst quiescence, see e.g. \cite{2015ApJ...804...87L}.

Thus, we expect that Model~\ref{m:bb} should well describe NUV-NIR observations at the beginning of XN outburst (thermal emission from almost isothermal disc ring), and Model~\ref{m:disc} should appear closer to the outburst maximum, when the automodel solution with constant mass accretion rate along the radius in the outer disc is established. As we will show in the section \S\ref{sec:analysis_discussion}, the observed spectral evolution during outburst rise in Aql~X-1 (SED measurements  \ref{sed:1}, \ref{sed:2} and \ref{sed:4}) qualitatively agrees with this theoretical picture. Below we describe the chosen spectral models and their parameters in detail.
  
\paragraph*{Model~(A).}
The adopted $bbodyrad$ blackbody model in XSPEC has two parameters: temperature $T_{bb}$ and normalisation $K_{bb}$. The normalisation parameter is connected to the projected emitting area $S_{bb}$ [cm$^2$] in the following way:
\begin{equation}
\label{eq:Sbb}
S_{bb}=\frac{\pi{}D_{5}^2}{4\cdot10^{-10}}\times{}K_{bb} ~,
\end{equation}
where $D_5$ is the source distance in units [5\,kpc].

\paragraph*{Model~(B).}
In order to model the irradiated accretion disc SED we adopted the popular DISKIR\footnote{https://heasarc.gsfc.nasa.gov/xanadu/xspec/models/diskir.html} model in XSPEC (without the inner disc coronal emission component, see Appendix~\ref{sec:A-diskir} for details). The adopted model has 3 parameters: $T_{\rm in,keV}$, $logrout$ and $f_{\rm out}$ (if irradiation is turned off --- $f_{\rm out}=0$). For a given X-ray luminosity illuminating the disc, these DISKIR parameters can be readily converted (see Appendix~\ref{sec:A-diskir}) into physical parameters of the outer accretion disc --- mass accretion rate $\dot{M}_{\rm out}$, disc outer radius $R_{\rm out}$ and the irradiation parameter $C$, which determines the fraction of X-ray flux thermalised in the disc. 

The disc outer radius in the Model~\ref{m:disc} can be constrained by the tidal truncation radius $R_{\rm out}<R_{\rm tid}$, which can be transformed into the following constraint on the DISKIR model parameter: 
\begin{equation}
logrout <  5.07 - \lg{D_5}
\end{equation}
for the adopted Aql~X-1 orbital parameters (see \S\ref{sec:aqlx1}) and using equation~(\ref{eq:ARout}).
 By taking into account uncertainty in the source distance, we get the following constraint: $logrout<5.15$. 
 No other constraints on the disc model parameters were applied during the fitting procedure. 

In the multi-color accretion disc Model~\ref{m:disc}, the irradiation parameter $C$ determines the degree of disc heating by X-ray irradiation. The surface temperature at outer radii of the accretion disc model (see Appendix~\ref{sec:A-diskir}) can be expressed as  
\begin{equation}
\sigma T^4 = \frac{3 G M_1 \dot{M}}{8\pi R^3} + C\frac{L_{\rm X}}{4\pi R^2} ~.
\label{eq:sigmaT4}
\end{equation}
The irradiation parameter contains information about geometry of the irradiated disc surface (disc height radial profile $H(R)$, disc albedo $a_{\rm out}$ and thermalisation fraction $\eta_{th}$ for X-ray photons:
\begin{equation}
C = \bigg(\frac{d{}H}{d{}R}-\frac{H}{R}\bigg)(1-a_{\rm out})\eta_{th} ~.
\end{equation} 
Assuming that the effective disc height, which intercepts X-rays (it can be the height of hot atmosphere or wind outflow formed above the outer disc, rather that photospheric disc height - see \cite{2002ApJ...581.1297J}) is a power law function of radius $H\propto{}R^{n}$ (e.g. $n=9/8$ for the outer zone of standard Shakura-Sunyaev accretion disc \citep{1973AA....24..337S}, $n=9/7$ for isothermal disc model of \cite{1990AA...235..162V}), one can obtain the following expression:
\begin{equation}
\label{eq:A-Ctheory}
C = (n-1)\frac{H}{R}(1-a_{\rm out})\eta_{th} ~.
\end{equation}
Note, that in the adopted DISKIR model the limiting case $H/R=const$ is assumed. In reality, $\frac{H}{R}\propto{}R^{n-1}$ is expected to be a slow function of radius $n-1=\frac{1}{8}\div\frac{2}{7}$ for the stationary accretion disc and the exact form of disc height radial profile $H(R)$ make sense for FUV part of disc spectrum. The far UV spectral range $\lambda<2000$\AA{}  is difficult to observe (the FUV observational data are currently absent for most X-ray Novae) and is very model-dependent to fit due to a strong extinction in this spectral range. We conclude, that for NIR-Optical-NUV spectral range the DISKIR model with $C=const_{R}$ seems to be an adequate choice for steady-state irradiated accretion disc model.        

\subsection{"X-ray tomograph" at the moment of hard/soft X-ray state transition in Aql~X-1}
\label{sec:xray_tomograph}
 The most remarkable moment at the Aql~X-1 outburst rise light curve is the hard/soft X-ray state transition (luckily covered by our SED measurements \ref{sed:2}--\ref{sed:3}). During the short time interval $\Delta{}T_{h/s}$, the fast changes in the structure of the inner accretion flow (optically-thin geometrically thick RIAF $\rightarrow$ optically-thick geometrically-thin standard Shakura-Syunyaev disc) are accompanied by a drastic softening of the X-ray spectrum: the amount of X-ray photons with $E>10$\,keV radically goes down ($f_{bol}$:$1.96\rightarrow1.08$). The fast evolution of the X-ray spectrum (heating the outer accretion disc) at the moment of state transition, can serve as "X-ray tomograph" to reveal the vertical structure and energy-depended X-ray heating efficiency of the outer accretion flow in X-ray Novae. 
 
With a reasonable assumption, that duration of X-ray state transition interval $\Delta{}T_{h/s}$ is short with respect to viscous time scale at the outer radii of accretion disc, the mass distribution (surface density radial profile $\Sigma_{\rm out}(R)$, accretion rate radial profile $\dot{M}_{\rm out}(R)$) in the outer disc should experience only a minimal changes during interval $\Delta{}T_{h/s}$. Indeed, for the standard \citep{1973AA....24..337S} accretion disc, the mass distribution at radius $R$ changes at a viscous timescale
\begin{equation}
\tau_{vis} = \frac{2}{3\alpha}\bigg(\frac{H}{R}\bigg)^{-2}\frac{1}{\Omega_K(R_d)} ~,
\end{equation}
where $\Omega_K=\sqrt{G{}M_1/R^3}$ is the Keplerian frequency, $M_1$ is the mass of the primary and $\alpha$ is the dimensionless viscosity parameter (see e.g. \cite{2005astro.ph..1215G}, \cite{2002apa..book.....F}). By adopting $\alpha=0.2$ \citep{2007MNRAS.376.1740K}, $H/R=0.1-0.2$ \citep{1996AA...314..484D,2002ApJ...581.1297J} and common neutron star mass $M_1=1.4\,M_\odot$ we obtain $\tau_{vis}=3.5-14^d\gg\Delta{}T_{h/s}$ for the Aql~X-1 outer disc radius estimate $R_d=R_{\rm tid}$. 

On the other hand, the temperature structure in the photosphere layers of the outer disc, which emit the observed NUV-Optical-NIR spectrum, can change substantially during the state transition interval, as it is directly governed by X-ray illumination (reprocessing time in the disc and its hot atmosphere for X-ray photons $\tau_{repr}\ll{}\Delta{}T_{h/s}$, see $\tau_{repr}$ estimates in \cite{1987ApJ...315..162C,2011AstL...37..826M}). By using SED measurements \ref{sed:2}--\ref{sed:3} at the edges of hard/soft X-ray state transition interval, we can test the vertical structure and energy-depended X-ray heating efficiency of the outer accretion disc in X-ray Novae Aql~X-1.

In this work, the main observable in the multi-color disc spectral model, which will be tested (see \S\ref{sec:analysis_discussion}) against various regimes of energy-depended X-ray heating, is the disc irradiation parameter $C$. Other parameters of the Model~B ($\dot{M}_{\rm out}$ and $R_{\rm out}$) are expected not to vary during the X-ray state transition interval, if the condition $\Delta{}T_{h/s}\ll{}\tau_{vis}$ is satisfied (see above).  
We will consider three qualitative choices for energy-dependent X-ray heating of the outer accretion disc:
\begin{itemize}   
\item heating by bolometric X-ray flux ($0.5-100$\,keV),
\item heating by soft X-rays ($0.5-10$\,keV).
\item heating by hard X-rays ($10-100$\,keV).
\end{itemize}
Consequently, in addition to irradiation parameter $C$ (which corresponds to "bolometric" X-ray heating, see above), it is straightforward to consider also the "soft" and "hard" irradiation parameters $C_s$ and $C_h$ with respect to $0.5-10$~keV and $10-100$~keV flux, respectively. The determination of "soft" irradiation parameter is justified in the case of direct illumination of the outer accretion disc by X-ray photons from the central source. Then soft X-rays with energies $\approx2\div10$\,keV may play a primary role in the heating of the outer disc surface (see e.g. \cite{1999AA...350...63S}). On the other hand, if direct illumination of the disc is not possible for some reason (e.g. due to concave disc height profile $H\propto{}R^{<1}$ or disc self-screening effect, see \cite{1999MNRAS.303..139D}), then the  hard ($E\gtrsim10$\,keV) X-rays, effectively scattered in the optically thin layers above the disc, may play a substantial role in the disc heating \citep{2011AstL...37..311M}.  It is worth noting, that  $C_s$, $C_h$ and $C$ parameters are simply connected to each other by using bolometric correction coefficient ($f_{bol}$):
\begin{eqnarray}\nonumber
C_{s} &=& C\times{}f_{bol} ~, \\
C_{h} &=& C\times\frac{f_{bol}}{f_{bol}-1} ~.
\label{eq:C-soft-hard}
\end{eqnarray}
For the Aql~X-1 outburst we adopt approximate bolometric corrections $f^{hard}_{bol}=1.96$ and $f^{soft}_{bol}=1.08$ for time moments before and after X-ray state transition, respectively (see \S\ref{sec:results}).


  
\section{Results}
\label{sec:analysis_discussion}
Four broad-band SED measurements \ref{sed:1}--\ref{sed:4} were obtained during the Aql~X-1 outburst rise phase (see \S\ref{sec:broadband-SED-points}) and were fitted by two spectral models \ref{m:bb} and \ref{m:disc}, described in the section \S\ref{sec:broadband-SED-models} above. The best fit parameters for black-body Model~\ref{m:bb} and multi-color disc Model~\ref{m:disc} with ($f_{\rm out}>0$) and without irradiation ($f_{\rm out}=0$) are presented in the Table~\ref{tbl:bestfit}. As can be noted, the inclusion of X-ray irradiation substantially improves multi-color disc fit for SED \ref{sed:1}, \ref{sed:2} and \ref{sed:4}. The first three columns in the Table~\ref{tbl:bestfit} contain: time $t$ with respect to state transition midpoint, orbital phase $\phi$ calculated from Aql~X-1 ephemeris (see Table~\ref{tbl:aqlx1_pars}) and bolometric X-ray luminosity in Eddington units calculated in the following way:
$\frac{L_{\rm X,bol}}{L_{\rm Edd}} = \frac{4\pi D^2 F_{\rm X,0.5-10} f_{bol}}{1.75\cdot10^{38} erg/s}$, where the value of Eddington limit is taken for pure hydrogen composition and $1.4{}M_\odot$ NS.


All best fit spectral models, together with SED data points, are shown at Figures~\ref{fig:sed:124} and \ref{fig:sed:3} by solid (blackbody), dashed (multi-color disc) and long dashed (multi-color disc with irradiation) lines. Both absorbed and unabsorbed curves for each model are shown (at Figure~\ref{fig:sed:3} only one unabsorbed curve is shown for clarity). All spectral curves are smoothed with top-hat window $\Delta\lambda/\lambda=0.25$ for better visual comparison with SED measurements, obtained in the broadband filters having relative bandwidth in the range $\Delta\lambda/\lambda=0.14\div0.34$ (see Table~\ref{tbl:ph_sys}). The smoothing is primary important on absorbed model curves in the NUV range, where model flux changes sharply with $\nu$. We note, that we derived smoothed model curves only for visualization purposes at Figures~\ref{fig:sed:124}-\ref{fig:sed:3}; all $\chi^2_r$ values presented in Table~\ref{tbl:bestfit} were obtained in the XSPEC fitting framework.  Below we discuss the derived results in detail.

Firstly, we consider SED measurements carried out during the Aql~X-1 outburst rise in the hard X-ray state --- \ref{sed:1}, \ref{sed:2} and SED obtained near the outburst maximum \ref{sed:4}. A curious spectral evolution during the state transition interval (SEDs \ref{sed:2}--\ref{sed:3}) will be discussed in the next section  \S\ref{sec:SED_evolution_at_state-transition}. 


A first SED \ref{sed:1} was obtained from combination of quasi-simultaneous \swiftuvot, \smarts and \rtt data (see \S\ref{sec:broadband-SED-points}) around a time moment $t=-6.05^d$. As can be seen from Table~\ref{tbl:bestfit} and Figure~\ref{fig:sed:124}~(left panel), the black-body model gives a substantially better fit than a multi-color disk model without irradiation. The best-fit blackbody model gives a goodness of the fit $\chi^2_r=1.01$ and the multi-color disc model gives $\chi^2_r=2.17$ (11 degrees of freedom), which rejects a later model with a p-value=0.013. With inclusion of X-ray irradiation, the multi-color disc model fit can be improved significantly. E.g. for irradiation parameter $C=2.9\cdot10^{-3}$ (see Table~\ref{tbl:bestfit}) the goodness of the fit reaches $\chi^2_r=1.27$ (11 d.o.f.), with the best-fit parameters $\dot{M}_{\rm out}\approx0.18{}\dot{M}_{\rm Edd}$, $R_{\rm out}\approx0.46{}R_{tidal}$. The fit can be further improved with increasing $C$. However, for the irradiation parameter value $C=2.9\cdot10^{-3}\gg\frac{3{}G{}M_1{}\dot{M}_{\rm out}}{2{}L_{\rm X}{}R_{\rm tid}}\approx3\cdot10^{-5}$, the optical flux from the disc is dominated by X-ray reprocessing (see equation (\ref{eq:Crepr})). Around the time moment $t=-6.05^d$, X-ray flux shows a fast changes $L_{\rm X}\propto{}e^{t/2^d}$ (see Figure~\ref{fig:lcurve_Xray}). For the X-ray reprocessing mechanism, one may expect the NIR-Optical/X-ray flux correlation in the broad-band filters $L_{Opt}\propto{}L_{\rm X}^{0.25-0.5}$ (see \cite{2012MNRAS.421.2846R,1994AA...290..133V}), which corresponds to the visual magnitude changes $\Delta{}m=-2.5\cdot\lg{\big(e^{0.125..0.25}\big)}\approx-0.14^{mag}..-0.27^{mag}$ per day. In contrary, the available \rtt observations  show almost constant g',r',i',z' fluxes around the time moment $t=-6.05^d$ (see Figure~\ref{fig:lcurve_opt}). Therefore, we may prefer the single temperature blackbody emission Model~\ref{m:bb} for the SED measurement \ref{sed:1}. The emitting disc ring should be heated primary by viscous dissipation (as no significant Optical/X-ray correlation is observed). The relative width of the blackbody emitting disc ring at radius $R$ can be estimated as 
\begin{equation}
\frac{\Delta{}R}{R}\approx\frac{S_{bb}}{\pi{}R^2\cos{(i)}} = \frac{K_{bb}}{4\cdot10^{-10} \cos{(i})}\bigg(\frac{D_5}{R}\bigg)^2 ~.
\end{equation}
(see equation~(\ref{eq:Sbb})). By supposing a disc ring is located at a tidal radius $R\approx{}R_{\rm tid}$ and adopting $D_5=1$, $i=42^\circ$ for Aql~X-1 (see Table~\ref{tbl:aqlx1_pars},) we get the following estimate: $\frac{\Delta{}R}{R}\approx0.15$. 

The next SED measurement \ref{sed:2} was obtained  just before hard/soft X-ray state transition at $t=-0.43^d$. Neither black-body ($\chi^2_r=11.7$, 4 d.o.f.), nor multi-color disc without irradiation ($\chi^2_r=11.0$, 4 d.o.f.) provide an acceptable fit for this SED measurement, but it can be well described ($\chi^2_r=0.43$, 3 d.o.f.) by the irradiated multi-color disc Model~\ref{m:disc} with reliable parameters: $\dot{M}_{\rm out}=0.66{}\dot{M}_{\rm Edd}$, $R_{\rm out}=0.94\cdot{}R_{tidal}$ and $C=6.1\times10^{-4}$. We adopt Eddington mass accretion rate value: $\dot{M}_{\rm Edd}=\frac{1.75\cdot10^{38}}{0.1 c^2}=1.95\cdot10^{18}$\,g/s \citep{2000MNRAS.314..498M}.

The final SED measurement \ref{sed:4} was carried out at a time moment $t=+1.8^d$ near the X-ray outburst maximum. As can be seen from Table~\ref{tbl:bestfit}, the multi-color disc without irradiation is statistically unacceptable model for this SED. Both irradiated multi-color disc or single-temperature  black-body models provide good fit to the SED data points. We note, that there are only 3 flux measurements ($J$, $R$, $U$ bands) combined in this SED, and NUV flux (M2/W2 bands at $\sim2000$\AA{}) measurement is not available at this time moment. We expect a degeneracy between physical parameters $\dot{M}$ and $C$ in the Model~\ref{m:disc}. Therefore we decided to fix a mass accretion rate during the fit, to the reasonable value estimated for the SED \ref{sed:2} (at $t=-0.46^d$). The Model~\ref{m:disc} with fixed $\dot{M}_{\rm out}=0.66\,\dot{M}_{\rm Edd}$ provides an acceptable fit with $\chi^2_r=0.03$ (1 d.o.f.) with a best-fit parameters $C=1.1\cdot10^{-3}$ and $R_{\rm out}\approx1.14{}R_{tidal}$(see Table~\ref{tbl:bestfit} and Figure~\ref{fig:sed:124} right panel). From numerical simulation of outbursts in X-ray Novae (see e.g. Figure~5 in \cite{2001AA...373..251D}) it is expected, that multi-color disc spectral energy distribution is already established at the moment of outburst maximum. Therefore, we also may prefer the irradiated multi-color disc as a best model for the SED \ref{sed:4}. 

In sum, we make a conclusion, that the observed SED evolution \ref{sed:1}, \ref{sed:2}, \ref{sed:4} during outburst rise in Aql~X-1 can be well understood as thermal emission from unstationary accretion disc flow with temperature radial distribution transforming from $\sim$ single-temperature blackbody emitting ring (heated primary by viscous dissipation) at early stages of outburst into the multi-color irradiated accretion disc measured around the X-ray outburst maximum. 


\begin{landscape}
\begin{table}
  \caption{Aql~X-1 SEDs best fits with $REDDEN*BBODYRAD$ (Model~A) and $REDDEN*DISKIR$ (Model~B) spectral models with fixed interstellar absorption $E_{B-V}=0.65^{mag}$. The reported errors on parameters of models correspond to 2$\sigma$ confidence level.} 
  \label{tbl:bestfit}
  \begin{tabular}{cccc c llcllllclllll}
    \hline
 \#SED&$t,$  &$\phi$ &$L_{\rm X,bol}$,&     {\bf A:} & $T_{bb}$,  & $K_{bb}$,        &$\chi_r^2$&       {\bf B:}  & $kT_{\rm in}$, & $logrout$        & $f_{\rm out}$,            & $\chi_r^2$ & $\dot{M}$,           & $R_{\rm out}$,& $C$,      & $C_s$,   & $C_h$, \\
           & $d$ &           & $L_{\rm Edd}$  &                 & eV             & $10^{11}$       &    (d.o.f.)   &                    & $keV$       &                        & $10^{-3}$            &  (d.o.f.)      & $\dot{M}_{\rm Edd}$ &  $R_{\rm tid}$ &$10^{-3}$&$10^{-3}$& $10^{-3}$ \\
\hline
(i)   & $-6.02$& $0.81$  &$0.017$   &         & $1.14\pm0.11$ & $8.00\pm1.94$& 1.01 (11)  &            & $1.91$               &$4.85$             &                             & 2.17 (11)  &  \\
      &             &              &                &         &                          &                         &                 &             & $1.49\pm0.07$ &$4.73\pm0.07$ & $0.1$                  & 1.27 (11)   & $0.183$              & $0.46$      & $2.9$ \\
\hline
(ii)  & $-0.46$& $0.90$  & $0.20$    &         & $1.10$              & $17.8$             &11.7 (4)     &             & $2.78$              & $5.15$            &                            & 10.4 (4)    & \\
      &             &              &                &         &                          &                         &                 &             & $2.05\pm0.31$ & $5.04\pm0.07$&$0.069$               & 0.43 (3)    & $0.66$                & $0.94$      & $0.61$        & $1.20$     &  $1.25$ \\
\hline
(iii) &$+0.55$& $0.13$  & $0.33$    &         &$0.80$               & $78.7$             & 2.76 (4)    &             & $3.19$               & $5.15$           &                            & 89.2 (4)    & \\
      &             &              &                &         &                          &                         &                 &             & $2.05$               & $5.04$            &$0.114$               &                  & $0.66$                & $0.94$      & $0.61$ \\
      &             &              &                &         &                          &                         &                 &             & $2.05$               & $5.04$            &$0.206$               &                  & $0.66$                & $0.94$      &                    & $1.20$ \\      
      &             &              &                &         &                          &                         &                 &             & $2.05$               & $5.04$            &$0.0172$             &                  & $0.66$                & $0.94$      &                     &               & $1.25$ \\

\hline
(iv) & $+1.80$& $0.71$ & $0.50$    &         & $1.07\pm0.03$ &$66\pm5$        & 0.38 (1)     &             & $3.96$              & $5.13$            &                            & 31.2 (1)    & \\	
      &              &             &                &         &                          &                        &                  &              & $2.05$ (fix)       & $5.12\pm0.04$& $0.315$             & 0.03 (1)    & $0.66$                 & $1.14$       & $1.11$ \\

\hline
  \end{tabular}
 \end{table}
 

 \begin{figure}  
  \includegraphics[width=7.0cm]{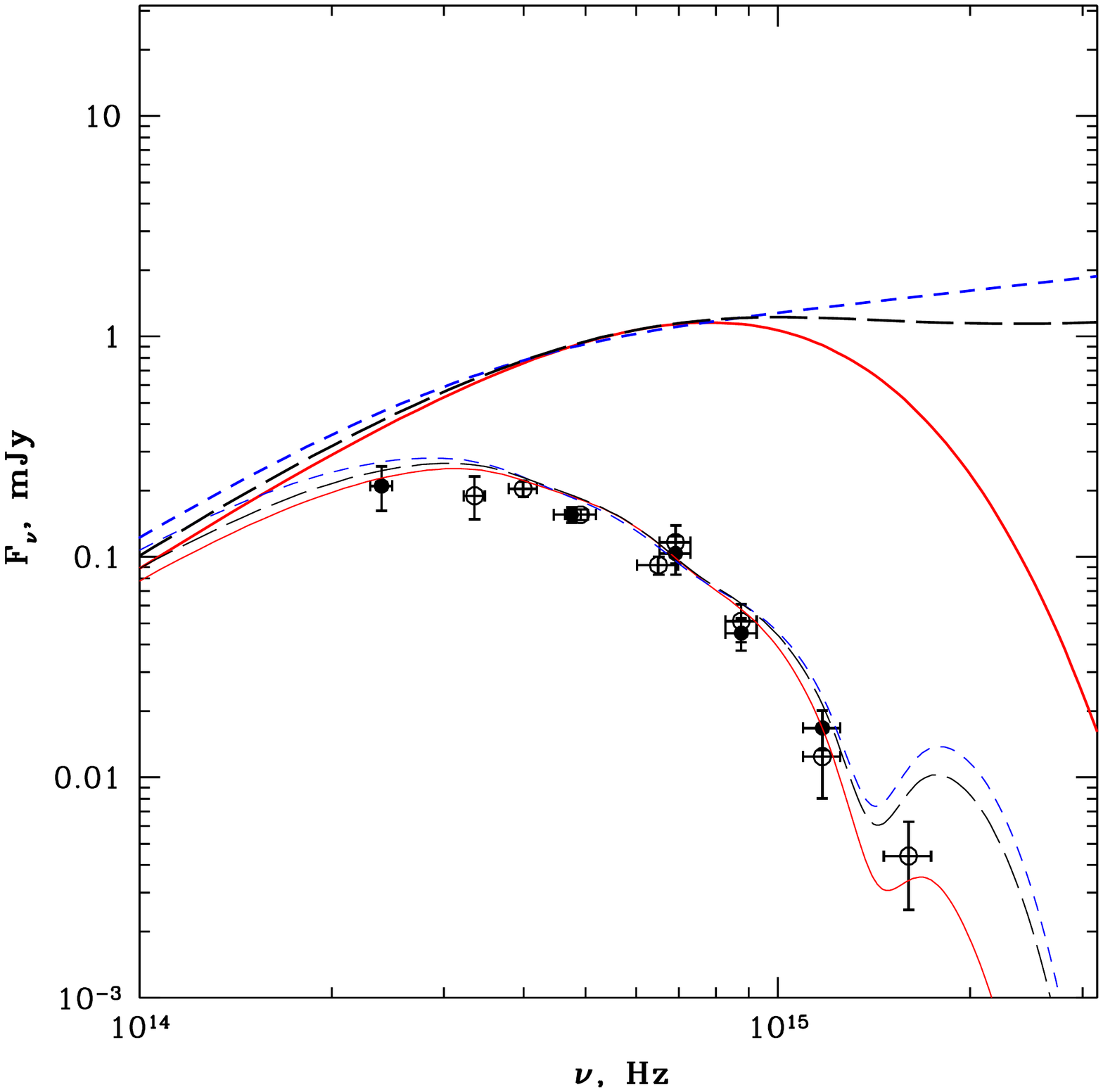}
  \includegraphics[width=7.0cm]{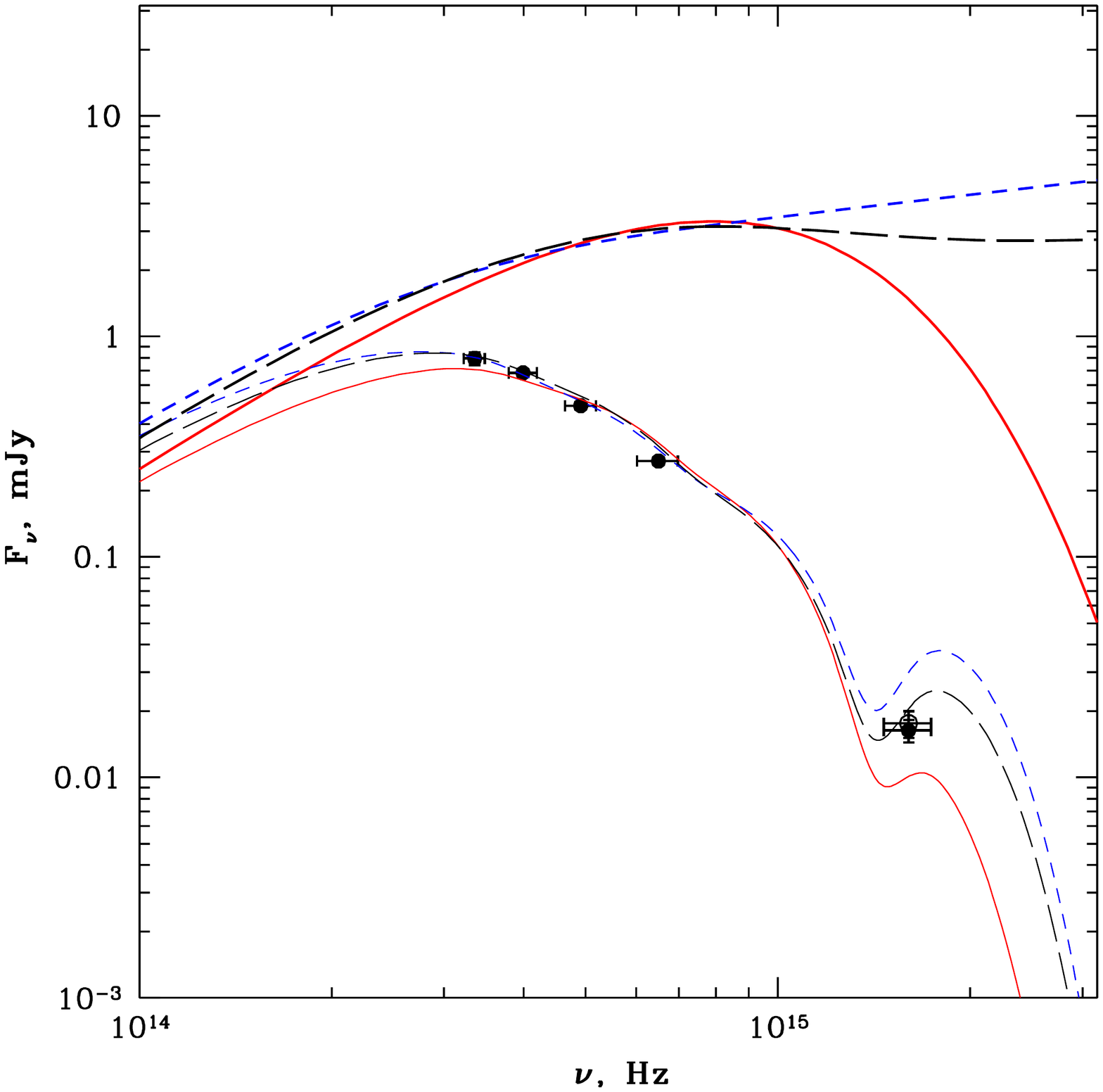}
  \includegraphics[width=7.0cm]{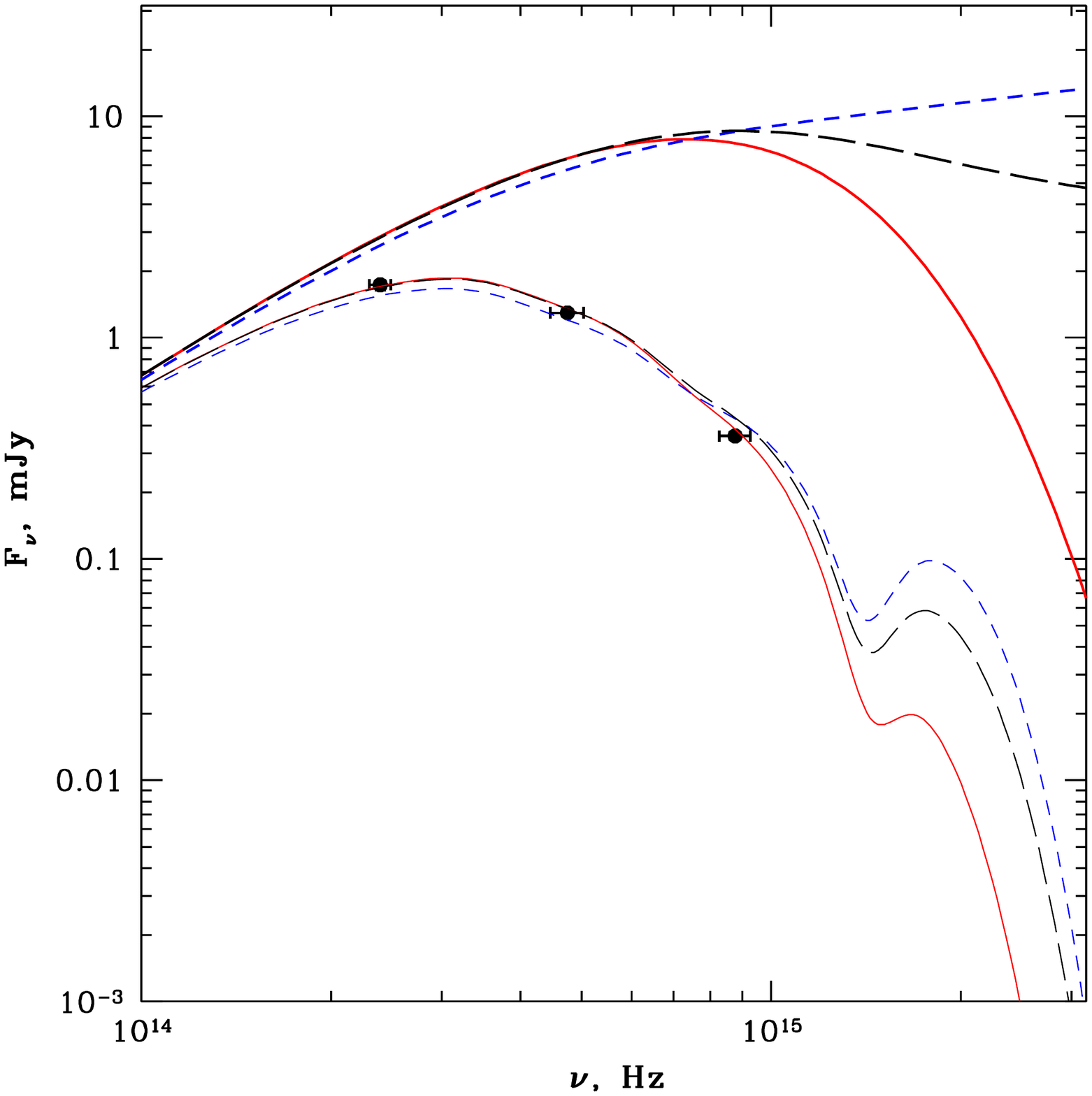}
    \caption{SED measurements \ref{sed:1} (left panel),\ref{sed:2} (central panel) and \ref{sed:4} (right panel), carried out at time moments $t=-6.02^d$, $-0.46^d$ and $+1.8^d$ (with respect the middle of X-ray state transition).}
  \label{fig:sed:124} 
  \end{figure}

\end{landscape}

  \begin{figure}  
  \includegraphics[width=8.5cm]{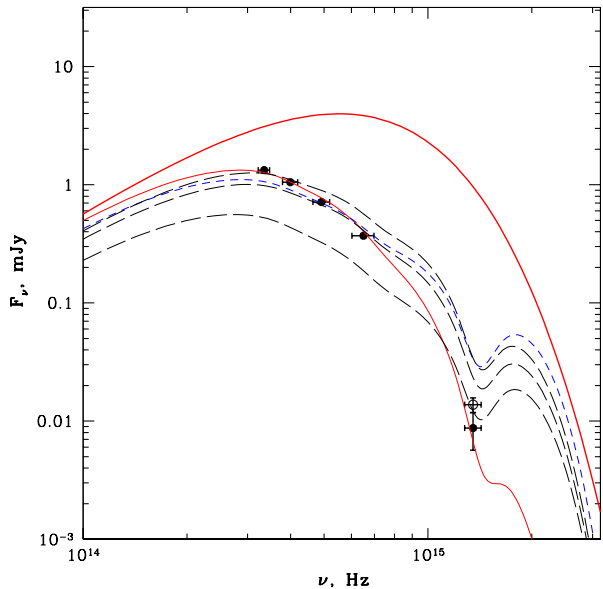}
    \caption{SED measurement \ref{sed:3} at the time moment $t=0.55^d$ (with respect the middle of X-ray state transition).}
  \label{fig:sed:3} 
  \end{figure}
  
\subsection{Evolution of the broadband SED during the hard/soft X-ray state transition in Aql X-1}
\label{sec:SED_evolution_at_state-transition}
The X-ray state transition interval during the outburst rise is covered by two SED measurements \ref{sed:2} and \ref{sed:3} at time moments $t=-0.46^d$ and $+0.55^d$, luckily carried out quasi-simultaneously (within interval $<0.05^d$) by \swiftuvot and \rtt telescopes (see \S\ref{sec:broadband-SED-points}). As it was discussed above in \S\ref{sec:analysis_discussion}, the SED \ref{sed:2} (at the start of state transition) can be well fitted by Model~\ref{m:disc} with reasonable physical parameters of the irradiated accretion disc: mass accretion rate $\dot{M}_{\rm out}=0.66\cdot\dot{M}_{\rm Edd}$, outer radius $R_{\rm out}=0.94\cdot{}R_{tidal}$ and irradiation parameter $C=6.1\cdot10^{-4}$ (see Table~\ref{tbl:bestfit}). By having in mind theoretical considerations  presented in the \S\ref{sec:xray_tomograph}, one may expect, that the same Model~\ref{m:disc} (with fixed $\dot{M}_{\rm out}$, $R_{\rm out}$ and irradiation parameter) is expected to match SED measurement \ref{sed:3} at the end of state transition.  Let's consider, what we see in reality. 

As can be seen from Table~\ref{tbl:bestfit}, at the time moment $t=+0.55^d$ the best-fit spectral model is a single-temperature black-body Model~\ref{m:bb}. Surprisingly, the multi-color disc Model~\ref{m:disc} (with or without irradiation) provides unacceptable fit to the data. Note, if we exclude the second \swiftuvot measurement (which was carried out not fully simultaneously with \rtt observation, see \S\ref{sec:broadband-SED-points}) from consideration, then the best-fit blackbody Model~\ref{m:bb} with $T_{bb}=0.764\pm0.014$ and $K_{bb}=(91\pm14)\cdot10^{11}$ ($2\sigma$ errors) became fully statistically acceptable with the goodness of the fit equal to $\chi^2_r=0.83$ (3 d.o.f.). One can conclude, that at the end of state transition the black-body like SED in the broad 2000--9000\AA{} spectral range is measured (it is shown by solid line at Figure~\ref{fig:sed:3}).

In order to better understand the NUV-NIR spectral evolution during X-ray state transition, we derived the expected SED evolution for the standard irradiated accretion disc Model~\ref{m:disc} with fixed parameters $\dot{M}=0.66\cdot\dot{M}_{\rm Edd}$, $R_{\rm out}=0.94\cdot{R_{\rm tid}}$ (values, measured at the start of X-ray state transition - see Table~\ref{tbl:bestfit}). It is worth noted, that X-ray spectrum changes drastically during the state transition, and the disc heating may depend either on hard, soft or bolometric X-ray flux. According to \S\ref{sec:xray_tomograph} we will consider three qualitative choices for X-ray heating of the outer disc:  
\begin{itemize}
\item Accretion disc can be sensitive to X-ray photons in the full energy range $0.5\div100$\,keV. In this case, we calculate Model~\ref{m:disc} with fixed "bolometric" irradiation parameter $C=6.1\cdot10^{-4}$ (see Table~\ref{tbl:bestfit}). According to formula~(\ref{eq:ACir3}), the $f_{\rm out}$ parameter in the DISKIR model should be properly rescaled as $f_{\rm out}\propto{}F_{\rm X,bol}$. The resulted spectral model is shown at Figure~\ref{fig:sed:3} by long dashed (middle) line.
\item Alternatively, accretion disc can be heated only by soft $0.5\div10$\,keV X-ray photons. In this case we fix the "soft" irradiation parameter $C_s=1.2\cdot10^{-3}$ (see equation~(\ref{eq:C-soft-hard})) and the DISKIR parameter $f_{\rm out}$ should be properly rescaled as  $f_{\rm out}\propto{}F_{\rm X,0.5-10}$. The resulted spectral model is shown at Figure~\ref{fig:sed:3} by long dashed (upper) line.
\item Accretion disc can be heated primary by hard $10\div100$\,keV X-ray photons. Then, we fix the "hard" irradiation parameter: $C_h=1.25\cdot10^{-3}$ (see equation~(\ref{eq:C-soft-hard})). The DISKIR parameter $f_{\rm out}$ should be properly rescaled as  $f_{\rm out}\propto{}F_{\rm X,10-100}$. The resulted spectral model is shown at Figure~\ref{fig:sed:3} by long dashed (lower) line.
\end{itemize}
All considered irradiated disc models with fixed $C_{bol}$, $C_s$, $C_h$ are  shown in the Table~\ref{tbl:bestfit}. One can conclude (see Figure~\ref{fig:sed:3}), that the observed SED \ref{sed:3}, measured immediately after X-ray state transition clearly disagree with expectations for irradiated accretion disc for any choice of single irradiation parameter. 

As can be seen at Figures~\ref{fig:lcurve_uvot} and \ref{fig:lcurve_opt}, during the interval of X-ray transition, the NUV flux at $\sim2000$\AA{}(W2/M2 band) decays slightly, but the optical blue g'-band (at $\sim4700$\AA{}) shows a small rise and the flux in the NIR z'-band ($\sim9000$\AA{}) rises more significantly. At Figure~\ref{fig:sed:3} one can see, that the observed flux evolution in the NIR z'-band can be closely described by disc model, irradiated by soft X-ray photons (fixed $C_s$) and flux evolution in the NUV M2-band can be described by disc irradiated by hard X-ray photons (fixed $C_h$). Note, that at the time moment $t=1.8^d$, Aql~X-1 brightness in all measured NUV-NIR filters increase significantly, with the biggest relative flux rise detected in the NUV. We conclude that, the $\sim1^d$ delayed rise of NUV brightness in the interval $t=0.55-1.8^d$, which makes the form of observed SED similar to the irradiated disc, is a curious observational fact and needs to be explained. 


We think, that the observed Aql~X-1 broad-band spectral evolution during the X-ray state transition can be understood if one considers a different mechanism of X-ray heating for NUV- and NIR-emitting regions of the disc. Let us suppose, that NUV-emitting regions in the disc are heated primary by scattered (in the hot corona or wind formed above the optically-thick accretion flow, see e.g. \cite{2011AstL...37..311M}) hard ($E>10$\,keV) X-ray photons, doe to the possible screening of  these regions from direct X-ray photons from the central source at the moment of X-ray state transition. One can see from Figure~\ref{fig:sed:3} (lower long-dashed line), that the observed decrease of NUV emission ($\sim2000$\AA{}) is well explained in the disc model, sensitive to hard X-ray photons (with constant $C_h$). On the other hand, NIR-emitting more outer regions in the disc can be heated by direct (mainly soft $2-10$\,keV, see e.g. \cite{1999AA...350...63S}) X-ray photons from the central source, and the rise of NIR brightness ($\sim8000$\AA{}) during the X-ray state transition is well explained in the disc model with constant "soft" irradiation parameter $C_s$ (upper long-dashed line at Figure~\ref{fig:sed:3}).

%
  
\section{Conclusions}
\label{sec:conclusions}
We studied a time evolution of the broad-band (NUV-Optical-NIR) spectral energy distribution (SED) in NS X-ray Nova Aql~X-1 during the rise phase of a bright FRED-type outburst in 2013. By using quasi-simultaneous observations from Swift orbital observatory and \rtt, \azt, \smarts 1-m class optical telescopes, we show that evolution of broad-band SED can be understood in the framework of thermal emission from unstationary accretion disc, which temperature radial distribution transforms from a single blackbody emitting ring at early stages of outburst into the standard multi-color irradiated accretion disc, with irradiation parameter $C\approx6\cdot10^{-4}$, measured at the end of hard X-ray state and near the outburst maximum. 

By using photometric observations, carried out luckily exactly at the edges of X-ray hard/soft state transition interval, we find an interesting effect: a decrease in NUV flux during this time interval, accompanied by a flux rise in NIR-Optical bands. The NUV flux decrease correlates with the hard X-rays $E>10$\,keV drop during the X-ray state transition, and the Optical-NIR flux rise correlates with the soft X-rays rise during the same time interval. In our interpretation, at the moment of X-ray state transition in Aql X-1 the UV-emitting parts of the accretion disc are screened from direct X-ray photons from the central source and heated primary by hard X-rays, effectively scattered in the hot corona or wind formed above the optically-thick accretion flow. At the same time, the outer and colder regions of accretion disc, emit in the Optical-NIR and are primary heated by direct X-ray illumination.        

 \vspace{5mm}
We point out that simultaneous multi-wavelength observations during the fast X-ray state transition interval in LMXBs provide an effective tool to directly test the energy-dependent X-ray heating efficiency, vertical structure and accretion flow geometry in the outer regions of accretion disc in X-ray Novae.  

\section*{Acknowledgements}

This research was supported by the Russian Scientific Foundation grant 14-12-00146. AM is deeply thankful to Mike Revnivtsev, Galja Lipunova, Konstantin Malanchev, Dmitry Karasev, Andy Semena for useful and fruitful discussions.

This research has made use of the \maxi data provided by RIKEN, JAXA and the MAXI team, and \swiftxrt and BAT data obtained from the High Energy Astrophysics Science Archive Research Center of NASA. AM is thankful to the Swift PI, Neil Gehrels, for accepting our requests for ToO observations of Aql~X-1 with Swift /XRT.

AM, I.Kh, IB thank to T{\"U}BITAK, IKI and KFU for partial supports in using RTT150 (Russian-Turkish 1.5-m telescope in Antalya), which made our optical monitoring program of Aql~X-1 possible. For the observational results from \rtt telescope presented in section \S\ref{sec:results}, AM acknowledges a partial support of the Russian Government Program of Competitive Growth of Kazan Federal University, I.Kh and IB acknowledge a partial support by RFBR and Government of Tatarstan under the project 15-42-02573. 

This paper has made use of publicly available up-to-date SMARTS optical/near-infrared light curves. We note, that the Yale SMARTS XRB team is supported by NSF grants 0407063 and 070707 to Charles Bailyn.

{\it Facilities:}  \swiftxrt , \swiftbat , \maxi , \rtt , \azt , \smarts

\appendix
\section{Measuring physical parameters of irradiated accretion disc by using DISKIR spectral model}
\label{sec:A-diskir}
  \cite{2009MNRAS.392.1106G} build a simple model for Optical/UV emission from the stationary $\dot{M}(R)=const$ multi-color disc self-irradiated by inner parts of the disc and coronal emission in black hole binaries. The model DISKIR became publicly available among other additive models in XSPEC package.We adopted this model to fit NUV-Optical-NIR SED of NS X-ray Nova Aql~X-1.
  
The DISKIR model has 9 parameters:
\begin{enumerate}
\item $T_{\rm in,keV}[keV]$, innermost temperature of the unilluminated disk in units [keV]
\item $\gamma$, asymptotic power-law photon index
\item $T_{e,keV}$, electron temperature (high energy rollover) in units [keV]
\item $L_{\rm c}/L_{\rm d}$, ratio of luminosity in the Compton tail to that of the unilluminated disk
\item $f_{\rm in}$, fraction of luminosity in the Compton tail which is thermalized in the inner disk
\item $r_{\rm irr}$, radius of the Compton illuminated disk in terms of the inner disk radius
\item $f_{\rm out}$, fraction of bolometric flux which is thermalized in the outer disk
\item $logrout$, log10 of the outer disk radius in terms of the inner disk radius
\item Normalization parameter (as in diskbb model):
\begin{equation}
K=4\cdot10^{-10}\bigg(\frac{R_{\rm in}}{D_5}\bigg)^2{}\cos(i) ~,
\label{eq:A-K}
\end{equation}
where $R_{\rm in}$ --- inner disc radius in [cm], $D_5$ --- distance in units 5\,kpc, $i$ --- system inclination.
\end{enumerate}   
Among all parameters of the model, we are interested only in four of them (which define disc SED in the NUV-NIR spectral range): $K$, $T_{\rm in,keV}$, $logrout$ and $f_{\rm out}$. Other parameters were fixed to their default values: $\gamma=1.7$, $kT_{e,keV}=100$, $f_{\rm in}=0.1$, $r_{\rm irr}=1.2$ and $L_{\rm c}/L_{\rm d}=0$ - irradiation of the inner disc and coronal emission are turned off. By using equations from \cite{2009MNRAS.392.1106G} (see their \S3) parameters $K$,$T_{\rm in,keV}$, $logrout$, $f_{\rm out}$) can be converted into the physical parameters we are interested.

\paragraph*{Inner disc radius ($R_{\rm in}$).}
The inner disc radius can be expressed from (\ref{eq:A-K}) as follows:
\begin{equation}\label{eq:A-Rin}
R_{\rm in} = 5\cdot10^{4}\Bigg[\frac{K}{\cos(i)}\Bigg]^{1/2} D_{5} ~[cm].
\end{equation}
Note, that the choice of normalization parameter $K$ (and $R_{\rm in}$ itself) is somewhat arbitrary as long as we are interested only in the outer disc emission. Hereafter we fixed normalization parameter to the value $K=400$, which corresponds to the inner disc radius:
\begin{equation}\label{eq:A-Rin1}
R_{\rm in} = 10^{6}\times\bigg(\frac{D_5}{\sqrt{\cos(i)}}\bigg) ~[cm].
\end{equation}

\paragraph*{Outer disc radius ($R_{\rm out}$).}
The outer disc radius can be expressed as $R_{\rm out}=10^{logrout}R_{\rm in}$, where $logrout$ is a parameter in DISKIR model. By using (\ref{eq:A-Rin1}) we have:
\begin{equation}\label{eq:ARout}
R_{\rm out} = 10^6\bigg(\frac{D_5}{\sqrt{\cos(i)}}\bigg) 10^{logrout} ~[cm].
\end{equation}

\paragraph*{Mass accretion rate in the outer disc ($\dot{M}_{\rm out}$).}
DISKIR is a model for stationary accretion disc (mass accretion rate is constant with radius $\dot{M}=const_R$). In DISKIR model, the temperature of unilluminated disk from $R_{\rm in}$ to $R_{\rm out}$ is described the following formula:
\begin{equation}\label{eq:A-Tdiskir}
T_{vis}(R) = T_{\rm in}\bigg(\frac{R}{R_{\rm in}}\bigg)^{-3/4} ~, 
\end{equation}
where $R_{\rm in}$ depends on normalization parameter $K$ (see formula (\ref{eq:A-Rin})) and inner radius temperature $T_{\rm in}$ (in units keV --- $T_{\rm in,keV}$). At the same time, photospheric temperature at the outer radii ($R>>R_{\rm in}$) of the unilluminated accretion disc can be expressed (see \cite{1973AA....24..337S}) as    
\begin{equation} \label{eq:A-Tvis}
\sigma_{\rm SB}{}T_{vis}^4 = \frac{3 G M\dot{M}_{\rm out}}{8\pi{}R^{3}} ~,
\end{equation}
where $G$ --- gravitation constant, $\sigma_{\rm SB}$ --- Stefan-Bolzman constant and $M$ --- compact object mass. By using equations (\ref{eq:A-Rin1}, \ref{eq:A-Tdiskir}, \ref{eq:A-Tvis}) we can connect the mass accretion rate in the outer parts of the disc with the $T_{\rm in,keV}$ parameter of the DISKIR model:  
\begin{equation}\label{eq:A-Mdoto}
\dot{M}_{\rm out} = 4.636\cdot10^{16}T_{\rm in,keV}^4\times\frac{D_5^3}{m_{1.4}(\cos(i))^{3/2}}  ~ [g/s],
\end{equation}
where $m_{1.4}$ --- compact object mass in units [$1.4\,M_\odot$].

\paragraph*{Irradiation parameter ($C$).}   
Let's consider the case, the outer parts of accretion disc are irradiated by the central source of X-ray luminosity $L_{\rm X}$ (see, e.g. \cite{2011AstL...37..311M}, \cite{2001AA...373..251D}, \cite{1973AA....24..337S}).  The temperature of the illuminated disc at radius $R>>R_{\rm in}$ can be defined as:
\begin{equation}\label{eq:A-Cir}
\sigma_{\rm SB}{}T^4 = \frac{3{}G{}M\dot{M}_{\rm out}}{8\pi{}R^{3}} + C\cdot{}\frac{L_{X}}{4\pi{}R^{2}} ~,
\end{equation}
where $C$ is a disk irradiation parameter. The characteristic irradiation parameter, then X-ray irradiation dominates the heating in the standard Shakura-Syunyaev disc at a given radius $R$,
can be  expressed as
\begin{equation}
C>\frac{3 G M_1\dot{M}_{\rm out}}{2 L_{\rm X} R} ~.
\label{eq:Crepr}
\end{equation}

Temperature at outer radii of the illuminated disc in the DISKIR model is expressed by the formula:
\begin{equation}
T^4(R) = T_{\rm in}^4 \Bigg[ \bigg(\frac{R}{R_{\rm in}}\bigg)^{-3} + f_{\rm out} \bigg(\frac{R}{R_{\rm in}}\bigg)^{-2} \Bigg]
\end{equation}
As can be noted, the $f_{\rm out}$ parameter in the DISKIR model corresponds to self-illumination of the accretion disc (outer part of the disc is irradiated by the disc luminosity $L_{\rm d}=4\pi\sigma_{\rm SB}{R_{\rm in}^2{}T_{\rm in}^4}$).
We would like to use a DISKIR model for a more general case, when the outer parts of the disc are illuminated by the central source of arbitrary X-ray luminosity. Then the irradiation parameter $C$ (from formula (\ref{eq:A-Cir}) above) is connected to DISKIR model parameter $f_{\rm out}$ in the following way:
\begin{equation}\nonumber
C = f_{\rm out} \cdot \frac{4\pi\sigma_{\rm SB}{}R_{\rm in}^2{T_{\rm in}^4}}{L_{\rm X}} ~.
\end{equation}

X-ray luminosity of the central source $L_{\rm X}$ can be expressed as:
\begin{equation}\label{eq:A-Lx}
L_{\rm X} = 4\pi\zeta{}D^2 F_{\rm X}  ~,
\end{equation}
where $F_{\rm X}$, $D$ correspond to the observed X-ray flux and distance to the source; $\zeta$ --- emission anisotropy factor ($\zeta=1$ for isotropic emission). By using (\ref{eq:A-Rin1}) and (\ref{eq:A-Lx}), finally we obtain:
\begin{equation}\label{eq:A-Cir2}
C = \frac{4.320\cdot10^{-9}}{\zeta\cos(i)}\times\frac{f_{\rm out}T_{\rm in,keV}^4}{F_{X}} ~.
\end{equation}
By using equation (\ref{eq:A-Mdoto}) we get:
\begin{equation}
\label{eq:ACir3}
C = 0.9318\cdot10^{-9}\times\frac{f_{\rm out}\dot{M}_{16}}{F_{X}}\times\frac{m_{1.4}\sqrt{\cos(i)}}{\zeta{}D_5^3} ~,
\end{equation}
where $\dot{M}_{16}$ corresponds to mass accretion rate in units [10$^{16}$\,g/s].

\bibliographystyle{mnras}
\bibliography{aqlx-1_2013_rise}

\begin{thebibliography}{}
\makeatletter
\relax
\def\mn@urlcharsother{\let\do\@makeother \do\$\do\&\do\#\do\^\do\_\do\%\do\~}
\def\mn@doi{\begingroup\mn@urlcharsother \@ifnextchar [ {\mn@doi@}
  {\mn@doi@[]}}
\def\mn@doi@[#1]#2{\def\@tempa{#1}\ifx\@tempa\@empty \href
  {http://dx.doi.org/#2} {doi:#2}\else \href {http://dx.doi.org/#2} {#1}\fi
  \endgroup}
\def\mn@eprint#1#2{\mn@eprint@#1:#2::\@nil}
\def\mn@eprint@arXiv#1{\href {http://arxiv.org/abs/#1} {{\tt arXiv:#1}}}
\def\mn@eprint@dblp#1{\href {http://dblp.uni-trier.de/rec/bibtex/#1.xml}
  {dblp:#1}}
\def\mn@eprint@#1:#2:#3:#4\@nil{\def\@tempa {#1}\def\@tempb {#2}\def\@tempc
  {#3}\ifx \@tempc \@empty \let \@tempc \@tempb \let \@tempb \@tempa \fi \ifx
  \@tempb \@empty \def\@tempb {arXiv}\fi \@ifundefined
  {mn@eprint@\@tempb}{\@tempb:\@tempc}{\expandafter \expandafter \csname
  mn@eprint@\@tempb\endcsname \expandafter{\@tempc}}}

\bibitem[\protect\citeauthoryear{{Arnaud}}{{Arnaud}}{1996}]{1996ASPC..101...17A}
{Arnaud} K.~A.,  1996, in {Jacoby} G.~H.,  {Barnes} J.,  eds,  Astronomical
  Society of the Pacific Conference Series Vol. 101, Astronomical Data Analysis
  Software and Systems V. p.~17

\bibitem[\protect\citeauthoryear{{Arnaud}}{{Arnaud}}{2010}]{2010HEAD...11.0905A}
{Arnaud} K.~A.,  2010, in AAS/High Energy Astrophysics Division \#11. p.~668

\bibitem[\protect\citeauthoryear{{Bessell}, {Castelli}  \& {Plez}}{{Bessell}
  et~al.}{1998}]{1998AA...333..231B}
{Bessell} M.~S.,  {Castelli} F.,   {Plez} B.,  1998, \aap, \href
  {http://adsabs.harvard.edu/abs/1998AA...333..231B} {333, 231}

\bibitem[\protect\citeauthoryear{{Bisnovatyi-Kogan} \&
  {Giovannelli}}{{Bisnovatyi-Kogan} \&
  {Giovannelli}}{2016}]{2016arXiv160507013B}
{Bisnovatyi-Kogan} G.~S.,  {Giovannelli} F.,  2016, preprint, \href
  {http://adsabs.harvard.edu/abs/2016arXiv160507013B} {} (\mn@eprint {arXiv}
  {1605.07013})

\bibitem[\protect\citeauthoryear{{Blackburn}}{{Blackburn}}{1995}]{1995ASPC...77..367B}
{Blackburn} J.~K.,  1995, in {Shaw} R.~A.,  {Payne} H.~E.,   {Hayes} J.~J.~E.,
  eds,  Astronomical Society of the Pacific Conference Series Vol. 77,
  Astronomical Data Analysis Software and Systems IV. p.~367

\bibitem[\protect\citeauthoryear{{Buxton}, {Bailyn}, {Capelo}, {Chatterjee},
  {Din{\c c}er}, {Kalemci}  \& {Tomsick}}{{Buxton}
  et~al.}{2012}]{2012AJ....143..130B}
{Buxton} M.~M.,  {Bailyn} C.~D.,  {Capelo} H.~L.,  {Chatterjee} R.,  {Din{\c
  c}er} T.,  {Kalemci} E.,   {Tomsick} J.~A.,  2012, \mn@doi [\aj]
  {10.1088/0004-6256/143/6/130}, \href
  {http://adsabs.harvard.edu/abs/2012AJ....143..130B} {143, 130}

\bibitem[\protect\citeauthoryear{{Campins}, {Rieke}  \& {Lebofsky}}{{Campins}
  et~al.}{1985}]{1985AJ.....90..896C}
{Campins} H.,  {Rieke} G.~H.,   {Lebofsky} M.~J.,  1985, \mn@doi [\aj]
  {10.1086/113799}, \href {http://adsabs.harvard.edu/abs/1985AJ.....90..896C}
  {90, 896}

\bibitem[\protect\citeauthoryear{{Cardelli}, {Clayton}  \& {Mathis}}{{Cardelli}
  et~al.}{1989}]{1989ApJ...345..245C}
{Cardelli} J.~A.,  {Clayton} G.~C.,   {Mathis} J.~S.,  1989, \mn@doi [\apj]
  {10.1086/167900}, \href {http://adsabs.harvard.edu/abs/1989ApJ...345..245C}
  {345, 245}

\bibitem[\protect\citeauthoryear{{Casella}, {Altamirano}, {Patruno}, {Wijnands}
   \& {van der Klis}}{{Casella} et~al.}{2008}]{2008ApJ...674L..41C}
{Casella} P.,  {Altamirano} D.,  {Patruno} A.,  {Wijnands} R.,   {van der Klis}
  M.,  2008, \mn@doi [\apjl] {10.1086/528982}, \href
  {http://adsabs.harvard.edu/abs/2008ApJ...674L..41C} {674, L41}

\bibitem[\protect\citeauthoryear{{Chen}, {Shrader}  \& {Livio}}{{Chen}
  et~al.}{1997}]{1997ApJ...491..312C}
{Chen} W.,  {Shrader} C.~R.,   {Livio} M.,  1997, \apj, \href
  {http://adsabs.harvard.edu/abs/1997ApJ...491..312C} {491, 312}

\bibitem[\protect\citeauthoryear{{Chevalier}, {Ilovaisky}, {Leisy}  \&
  {Patat}}{{Chevalier} et~al.}{1999}]{1999AA...347L..51C}
{Chevalier} C.,  {Ilovaisky} S.~A.,  {Leisy} P.,   {Patat} F.,  1999, \aap,
  \href {http://adsabs.harvard.edu/abs/1999A%26A...347L..51C} {347, L51}

\bibitem[\protect\citeauthoryear{{Cominsky}, {London}  \& {Klein}}{{Cominsky}
  et~al.}{1987}]{1987ApJ...315..162C}
{Cominsky} L.~R.,  {London} R.~A.,   {Klein} R.~I.,  1987, \mn@doi [\apj]
  {10.1086/165122}, \href {http://adsabs.harvard.edu/abs/1987ApJ...315..162C}
  {315, 162}

\bibitem[\protect\citeauthoryear{{Corral-Santana}, {Casares},
  {Mu{\~n}oz-Darias}, {Bauer}, {Mart{\'{\i}}nez-Pais}  \&
  {Russell}}{{Corral-Santana} et~al.}{2016}]{2016AA...587A..61C}
{Corral-Santana} J.~M.,  {Casares} J.,  {Mu{\~n}oz-Darias} T.,  {Bauer} F.~E.,
  {Mart{\'{\i}}nez-Pais} I.~G.,   {Russell} D.~M.,  2016, \mn@doi [\aap]
  {10.1051/0004-6361/201527130}, \href
  {http://adsabs.harvard.edu/abs/2016A\%26A...587A..61C} {587, A61}

\bibitem[\protect\citeauthoryear{{Degenaar} \& {Wijnands}}{{Degenaar} \&
  {Wijnands}}{2013}]{2013ATel.5117....1D}
{Degenaar} N.,  {Wijnands} R.,  2013, The Astronomer's Telegram, \href
  {http://adsabs.harvard.edu/abs/2013ATel.5117....1D} {5117}

\bibitem[\protect\citeauthoryear{{Degenaar} et~al.,}{{Degenaar}
  et~al.}{2014}]{2014ApJ...784..122D}
{Degenaar} N.,  et~al., 2014, \mn@doi [\apj] {10.1088/0004-637X/784/2/122},
  \href {http://adsabs.harvard.edu/abs/2014ApJ...784..122D} {784, 122}

\bibitem[\protect\citeauthoryear{{Dickey} \& {Lockman}}{{Dickey} \&
  {Lockman}}{1990}]{1990ARAA..28..215D}
{Dickey} J.~M.,  {Lockman} F.~J.,  1990, \mn@doi [\araa]
  {10.1146/annurev.aa.28.090190.001243}, \href
  {http://adsabs.harvard.edu/abs/1990ARA\%26A..28..215D} {28, 215}

\bibitem[\protect\citeauthoryear{{Dubus}, {Lasota}, {Hameury}  \&
  {Charles}}{{Dubus} et~al.}{1999}]{1999MNRAS.303..139D}
{Dubus} G.,  {Lasota} J.-P.,  {Hameury} J.-M.,   {Charles} P.,  1999, \mn@doi
  [\mnras] {10.1046/j.1365-8711.1999.02212.x}, \href
  {http://adsabs.harvard.edu/abs/1999MNRAS.303..139D} {303, 139}

\bibitem[\protect\citeauthoryear{{Dubus}, {Hameury}  \& {Lasota}}{{Dubus}
  et~al.}{2001}]{2001AA...373..251D}
{Dubus} G.,  {Hameury} J.-M.,   {Lasota} J.-P.,  2001, \mn@doi [\aap]
  {10.1051/0004-6361:20010632}, \href
  {http://adsabs.harvard.edu/abs/2001A%26A...373..251D} {373, 251}

\bibitem[\protect\citeauthoryear{{Eggleton}}{{Eggleton}}{1983}]{1983ApJ...268..368E}
{Eggleton} P.~P.,  1983, \mn@doi [\apj] {10.1086/160960}, \href
  {http://adsabs.harvard.edu/abs/1983ApJ...268..368E} {268, 368}

\bibitem[\protect\citeauthoryear{{Esin}, {Kuulkers}, {McClintock}  \&
  {Narayan}}{{Esin} et~al.}{2000}]{2000ApJ...532.1069E}
{Esin} A.~A.,  {Kuulkers} E.,  {McClintock} J.~E.,   {Narayan} R.,  2000,
  \mn@doi [\apj] {10.1086/308615}, \href
  {http://adsabs.harvard.edu/abs/2000ApJ...532.1069E} {532, 1069}

\bibitem[\protect\citeauthoryear{{Evans} et~al.,}{{Evans}
  et~al.}{2009}]{2009MNRAS.397.1177E}
{Evans} P.~A.,  et~al., 2009, \mn@doi [\mnras]
  {10.1111/j.1365-2966.2009.14913.x}, \href
  {http://adsabs.harvard.edu/abs/2009MNRAS.397.1177E} {397, 1177}

\bibitem[\protect\citeauthoryear{{Frank}, {King}  \& {Raine}}{{Frank}
  et~al.}{2002}]{2002apa..book.....F}
{Frank} J.,  {King} A.,   {Raine} D.~J.,  2002, {Accretion Power in
  Astrophysics: Third Edition}

\bibitem[\protect\citeauthoryear{{Friedman}, {Byram}  \& {Chubb}}{{Friedman}
  et~al.}{1967}]{1967Sci...156..374F}
{Friedman} H.,  {Byram} E.~T.,   {Chubb} T.~A.,  1967, \mn@doi [Science]
  {10.1126/science.156.3773.374}, \href
  {http://adsabs.harvard.edu/abs/1967Sci...156..374F} {156, 374}

\bibitem[\protect\citeauthoryear{{Fukugita}, {Ichikawa}, {Gunn}, {Doi},
  {Shimasaku}  \& {Schneider}}{{Fukugita} et~al.}{1996}]{1996AJ....111.1748F}
{Fukugita} M.,  {Ichikawa} T.,  {Gunn} J.~E.,  {Doi} M.,  {Shimasaku} K.,
  {Schneider} D.~P.,  1996, \mn@doi [\aj] {10.1086/117915}, \href
  {http://adsabs.harvard.edu/abs/1996AJ....111.1748F} {111, 1748}

\bibitem[\protect\citeauthoryear{{Galloway}, {Muno}, {Hartman}, {Psaltis}  \&
  {Chakrabarty}}{{Galloway} et~al.}{2008}]{2008ApJS..179..360G}
{Galloway} D.~K.,  {Muno} M.~P.,  {Hartman} J.~M.,  {Psaltis} D.,
  {Chakrabarty} D.,  2008, \mn@doi [\apjs] {10.1086/592044}, \href
  {http://adsabs.harvard.edu/abs/2008ApJS..179..360G} {179, 360}

\bibitem[\protect\citeauthoryear{{Gehrels} et~al.,}{{Gehrels}
  et~al.}{2004}]{2004ApJ...611.1005G}
{Gehrels} N.,  et~al., 2004, \mn@doi [\apj] {10.1086/422091}, \href
  {http://adsabs.harvard.edu/abs/2004ApJ...611.1005G} {611, 1005}

\bibitem[\protect\citeauthoryear{{Gierli{\'n}ski}, {Done}  \&
  {Page}}{{Gierli{\'n}ski} et~al.}{2009}]{2009MNRAS.392.1106G}
{Gierli{\'n}ski} M.,  {Done} C.,   {Page} K.,  2009, \mn@doi [\mnras]
  {10.1111/j.1365-2966.2008.14166.x}, \href
  {http://adsabs.harvard.edu/abs/2009MNRAS.392.1106G} {392, 1106}

\bibitem[\protect\citeauthoryear{{Gilfanov} \& {Arefiev}}{{Gilfanov} \&
  {Arefiev}}{2005}]{2005astro.ph..1215G}
{Gilfanov} M.,  {Arefiev} V.,  2005, ArXiv Astrophysics e-prints, \href
  {http://adsabs.harvard.edu/abs/2005astro.ph..1215G} {}

\bibitem[\protect\citeauthoryear{{Grebenev}, {Prosvetov}, {Burenin}, {Krivonos}
   \& {Mescheryakov}}{{Grebenev} et~al.}{2016}]{2016AstL...42...69G}
{Grebenev} S.~A.,  {Prosvetov} A.~V.,  {Burenin} R.~A.,  {Krivonos} R.~A.,
  {Mescheryakov} A.~V.,  2016, \mn@doi [Astronomy Letters]
  {10.1134/S1063773716020031}, \href
  {http://adsabs.harvard.edu/abs/2016AstL...42...69G} {42, 69}

\bibitem[\protect\citeauthoryear{{G{\"u}ng{\"o}r}, {G{\"u}ver}  \& {Ek{\c
  s}i}}{{G{\"u}ng{\"o}r} et~al.}{2014}]{2014MNRAS.439.2717G}
{G{\"u}ng{\"o}r} C.,  {G{\"u}ver} T.,   {Ek{\c s}i} K.~Y.,  2014, \mn@doi
  [\mnras] {10.1093/mnras/stu128}, \href
  {http://adsabs.harvard.edu/abs/2014MNRAS.439.2717G} {439, 2717}

\bibitem[\protect\citeauthoryear{{Hameury}, {Lasota}, {McClintock}  \&
  {Narayan}}{{Hameury} et~al.}{1997}]{1997ApJ...489..234H}
{Hameury} J.-M.,  {Lasota} J.-P.,  {McClintock} J.~E.,   {Narayan} R.,  1997,
  \mn@doi [\apj] {10.1086/304780}, \href
  {http://adsabs.harvard.edu/abs/1997ApJ...489..234H} {489, 234}

\bibitem[\protect\citeauthoryear{{Hynes} \& {Robinson}}{{Hynes} \&
  {Robinson}}{2012}]{2012ApJ...749....3H}
{Hynes} R.~I.,  {Robinson} E.~L.,  2012, \mn@doi [\apj]
  {10.1088/0004-637X/749/1/3}, \href
  {http://adsabs.harvard.edu/abs/2012ApJ...749....3H} {749, 3}

\bibitem[\protect\citeauthoryear{{Jimenez-Garate}, {Raymond}  \&
  {Liedahl}}{{Jimenez-Garate} et~al.}{2002}]{2002ApJ...581.1297J}
{Jimenez-Garate} M.~A.,  {Raymond} J.~C.,   {Liedahl} D.~A.,  2002, \mn@doi
  [\apj] {10.1086/344364}, \href
  {http://adsabs.harvard.edu/abs/2002ApJ...581.1297J} {581, 1297}

\bibitem[\protect\citeauthoryear{{Kalberla}, {Burton}, {Hartmann}, {Arnal},
  {Bajaja}, {Morras}  \& {P{\"o}ppel}}{{Kalberla}
  et~al.}{2005}]{2005AA...440..775K}
{Kalberla} P.~M.~W.,  {Burton} W.~B.,  {Hartmann} D.,  {Arnal} E.~M.,  {Bajaja}
  E.,  {Morras} R.,   {P{\"o}ppel} W.~G.~L.,  2005, \mn@doi [\aap]
  {10.1051/0004-6361:20041864}, \href
  {http://adsabs.harvard.edu/abs/2005A\%26A...440..775K} {440, 775}

\bibitem[\protect\citeauthoryear{{King}, {Pringle}  \& {Livio}}{{King}
  et~al.}{2007}]{2007MNRAS.376.1740K}
{King} A.~R.,  {Pringle} J.~E.,   {Livio} M.,  2007, \mn@doi [\mnras]
  {10.1111/j.1365-2966.2007.11556.x}, \href
  {http://adsabs.harvard.edu/abs/2007MNRAS.376.1740K} {376, 1740}

\bibitem[\protect\citeauthoryear{{Kiziltan}, {Kottas}, {De Yoreo}  \&
  {Thorsett}}{{Kiziltan} et~al.}{2013}]{2013ApJ...778...66K}
{Kiziltan} B.,  {Kottas} A.,  {De Yoreo} M.,   {Thorsett} S.~E.,  2013, \mn@doi
  [\apj] {10.1088/0004-637X/778/1/66}, \href
  {http://adsabs.harvard.edu/abs/2013ApJ...778...66K} {778, 66}

\bibitem[\protect\citeauthoryear{{Koyama} et~al.,}{{Koyama}
  et~al.}{1981}]{1981ApJ...247L..27K}
{Koyama} K.,  et~al., 1981, \mn@doi [\apjl] {10.1086/183582}, \href
  {http://adsabs.harvard.edu/abs/1981ApJ...247L..27K} {247, L27}

\bibitem[\protect\citeauthoryear{{Landolt}}{{Landolt}}{1992}]{1992AJ....104..340L}
{Landolt} A.~U.,  1992, \mn@doi [\aj] {10.1086/116242}, \href
  {http://adsabs.harvard.edu/abs/1992AJ....104..340L} {104, 340}

\bibitem[\protect\citeauthoryear{{Lasota}}{{Lasota}}{2001}]{2001NewAR..45..449L}
{Lasota} J.-P.,  2001, \mn@doi [\nar] {10.1016/S1387-6473(01)00112-9}, \href
  {http://adsabs.harvard.edu/abs/2001NewAR..45..449L} {45, 449}

\bibitem[\protect\citeauthoryear{{Lipunova}}{{Lipunova}}{2015}]{2015ApJ...804...87L}
{Lipunova} G.~V.,  2015, \mn@doi [\apj] {10.1088/0004-637X/804/2/87}, \href
  {http://adsabs.harvard.edu/abs/2015ApJ...804...87L} {804, 87}

\bibitem[\protect\citeauthoryear{{Lipunova} \& {Malanchev}}{{Lipunova} \&
  {Malanchev}}{2016}]{2016arXiv161001399L}
{Lipunova} G.~V.,  {Malanchev} K.~L.,  2016, preprint, \href
  {http://adsabs.harvard.edu/abs/2016arXiv161001399L} {} (\mn@eprint {arXiv}
  {1610.01399})

\bibitem[\protect\citeauthoryear{{Lubow} \& {Shu}}{{Lubow} \&
  {Shu}}{1975}]{1975ApJ...198..383L}
{Lubow} S.~H.,  {Shu} F.~H.,  1975, \mn@doi [\apj] {10.1086/153614}, \href
  {http://adsabs.harvard.edu/abs/1975ApJ...198..383L} {198, 383}

\bibitem[\protect\citeauthoryear{{Lyubarskij} \& {Shakura}}{{Lyubarskij} \&
  {Shakura}}{1987}]{1987SvAL...13..386L}
{Lyubarskij} Y.~E.,  {Shakura} N.~I.,  1987, Soviet Astronomy Letters, \href
  {http://adsabs.harvard.edu/abs/1987SvAL...13..386L} {13, 386}

\bibitem[\protect\citeauthoryear{{Maitra} \& {Bailyn}}{{Maitra} \&
  {Bailyn}}{2008}]{2008ApJ...688..537M}
{Maitra} D.,  {Bailyn} C.~D.,  2008, \mn@doi [\apj] {10.1086/592029}, \href
  {http://adsabs.harvard.edu/abs/2008ApJ...688..537M} {688, 537}

\bibitem[\protect\citeauthoryear{{Mata S{\'a}nchez}, {Mu{\~n}oz-Darias},
  {Casares}  \& {Jim{\'e}nez-Ibarra}}{{Mata S{\'a}nchez}
  et~al.}{2016}]{2016arXiv160900392M}
{Mata S{\'a}nchez} D.,  {Mu{\~n}oz-Darias} T.,  {Casares} J.,
  {Jim{\'e}nez-Ibarra} F.,  2016, preprint, \href
  {http://adsabs.harvard.edu/abs/2016arXiv160900392M} {} (\mn@eprint {arXiv}
  {1609.00392})

\bibitem[\protect\citeauthoryear{{Menou}, {Hameury}, {Lasota}  \&
  {Narayan}}{{Menou} et~al.}{2000}]{2000MNRAS.314..498M}
{Menou} K.,  {Hameury} J.-M.,  {Lasota} J.-P.,   {Narayan} R.,  2000, \mn@doi
  [\mnras] {10.1046/j.1365-8711.2000.03357.x}, \href
  {http://adsabs.harvard.edu/abs/2000MNRAS.314..498M} {314, 498}

\bibitem[\protect\citeauthoryear{{Mescheryakov}, {Shakura}  \&
  {Suleimanov}}{{Mescheryakov} et~al.}{2011a}]{2011AstL...37..311M}
{Mescheryakov} A.~V.,  {Shakura} N.~I.,   {Suleimanov} V.~F.,  2011a, \mn@doi
  [Astronomy Letters] {10.1134/S1063773711050045}, \href
  {http://adsabs.harvard.edu/abs/2011AstL...37..311M} {37, 311}

\bibitem[\protect\citeauthoryear{{Mescheryakov}, {Revnivtsev}  \&
  {Filippova}}{{Mescheryakov} et~al.}{2011b}]{2011AstL...37..826M}
{Mescheryakov} A.~V.,  {Revnivtsev} M.~G.,   {Filippova} E.~V.,  2011b, \mn@doi
  [Astronomy Letters] {10.1134/S1063773711120073}, \href
  {http://adsabs.harvard.edu/abs/2011AstL...37..826M} {37, 826}

\bibitem[\protect\citeauthoryear{{Meshcheryakov} et~al.,}{{Meshcheryakov}
  et~al.}{2013}]{2013ATel.5114....1M}
{Meshcheryakov} A.,  et~al., 2013, The Astronomer's Telegram, \href
  {http://adsabs.harvard.edu/abs/2013ATel.5114....1M} {5114}

\bibitem[\protect\citeauthoryear{{Migliari} \& {Fender}}{{Migliari} \&
  {Fender}}{2006}]{2006MNRAS.366...79M}
{Migliari} S.,  {Fender} R.~P.,  2006, \mn@doi [\mnras]
  {10.1111/j.1365-2966.2005.09777.x}, \href
  {http://adsabs.harvard.edu/abs/2006MNRAS.366...79M} {366, 79}

\bibitem[\protect\citeauthoryear{{Miller-Jones} et~al.,}{{Miller-Jones}
  et~al.}{2010}]{2010ApJ...716L.109M}
{Miller-Jones} J.~C.~A.,  et~al., 2010, \mn@doi [\apjl]
  {10.1088/2041-8205/716/2/L109}, \href
  {http://adsabs.harvard.edu/abs/2010ApJ...716L.109M} {716, L109}

\bibitem[\protect\citeauthoryear{{Nakahira}, {Negoro}, {Shidatsu}, {Ueda},
  {Mihara}, {Sugizaki}, {Matsuoka}  \& {Onodera}}{{Nakahira}
  et~al.}{2014}]{2014PASJ...66...84N}
{Nakahira} S.,  {Negoro} H.,  {Shidatsu} M.,  {Ueda} Y.,  {Mihara} T.,
  {Sugizaki} M.,  {Matsuoka} M.,   {Onodera} T.,  2014, \mn@doi [\pasj]
  {10.1093/pasj/psu060}, \href
  {http://adsabs.harvard.edu/abs/2014PASJ...66...84N} {66, 84}

\bibitem[\protect\citeauthoryear{{Paczynski}}{{Paczynski}}{1977}]{1977ApJ...216..822P}
{Paczynski} B.,  1977, \mn@doi [\apj] {10.1086/155526}, \href
  {http://adsabs.harvard.edu/abs/1977ApJ...216..822P} {216, 822}

\bibitem[\protect\citeauthoryear{{Poole} et~al.,}{{Poole}
  et~al.}{2008}]{2008MNRAS.383..627P}
{Poole} T.~S.,  et~al., 2008, \mn@doi [\mnras]
  {10.1111/j.1365-2966.2007.12563.x}, \href
  {http://adsabs.harvard.edu/abs/2008MNRAS.383..627P} {383, 627}

\bibitem[\protect\citeauthoryear{{Predehl} \& {Schmitt}}{{Predehl} \&
  {Schmitt}}{1995}]{1995AA...293..889P}
{Predehl} P.,  {Schmitt} J.~H.~M.~M.,  1995, \aap, \href
  {http://adsabs.harvard.edu/abs/1995A\%26A...293..889P} {293}

\bibitem[\protect\citeauthoryear{{Reig}, {M{\'e}ndez}, {van der Klis}  \&
  {Ford}}{{Reig} et~al.}{2000}]{2000ApJ...530..916R}
{Reig} P.,  {M{\'e}ndez} M.,  {van der Klis} M.,   {Ford} E.~C.,  2000, \mn@doi
  [\apj] {10.1086/308399}, \href
  {http://adsabs.harvard.edu/abs/2000ApJ...530..916R} {530, 916}

\bibitem[\protect\citeauthoryear{{Revnivtsev}, {Zolotukhin}  \&
  {Meshcheryakov}}{{Revnivtsev} et~al.}{2012}]{2012MNRAS.421.2846R}
{Revnivtsev} M.~G.,  {Zolotukhin} I.~Y.,   {Meshcheryakov} A.~V.,  2012,
  \mn@doi [\mnras] {10.1111/j.1365-2966.2012.20511.x}, \href
  {http://adsabs.harvard.edu/abs/2012MNRAS.421.2846R} {421, 2846}

\bibitem[\protect\citeauthoryear{{Sakurai}, {Yamada}, {Torii}, {Noda},
  {Nakazawa}, {Makishima}  \& {Takahashi}}{{Sakurai}
  et~al.}{2012}]{2012PASJ...64...72S}
{Sakurai} S.,  {Yamada} S.,  {Torii} S.,  {Noda} H.,  {Nakazawa} K.,
  {Makishima} K.,   {Takahashi} H.,  2012, \mn@doi [\pasj]
  {10.1093/pasj/64.4.72}, \href
  {http://adsabs.harvard.edu/abs/2012PASJ...64...72S} {64}

\bibitem[\protect\citeauthoryear{{Schlafly} \& {Finkbeiner}}{{Schlafly} \&
  {Finkbeiner}}{2011}]{2011ApJ...737..103S}
{Schlafly} E.~F.,  {Finkbeiner} D.~P.,  2011, \mn@doi [\apj]
  {10.1088/0004-637X/737/2/103}, \href
  {http://adsabs.harvard.edu/abs/2011ApJ...737..103S} {737, 103}

\bibitem[\protect\citeauthoryear{{Schlegel}, {Finkbeiner}  \&
  {Davis}}{{Schlegel} et~al.}{1998}]{1998ApJ...500..525S}
{Schlegel} D.~J.,  {Finkbeiner} D.~P.,   {Davis} M.,  1998, \mn@doi [\apj]
  {10.1086/305772}, \href {http://adsabs.harvard.edu/abs/1998ApJ...500..525S}
  {500, 525}

\bibitem[\protect\citeauthoryear{{Shahbaz}, {Bandyopadhyay}, {Charles},
  {Wagner}, {Muhli}, {Hakala}, {Casares}  \& {Greenhill}}{{Shahbaz}
  et~al.}{1998}]{1998MNRAS.300.1035S}
{Shahbaz} T.,  {Bandyopadhyay} R.~M.,  {Charles} P.~A.,  {Wagner} R.~M.,
  {Muhli} P.,  {Hakala} P.,  {Casares} J.,   {Greenhill} J.,  1998, \mn@doi
  [\mnras] {10.1046/j.1365-8711.1998.01965.x}, \href
  {http://adsabs.harvard.edu/abs/1998MNRAS.300.1035S} {300, 1035}

\bibitem[\protect\citeauthoryear{{Shakura} \& {Sunyaev}}{{Shakura} \&
  {Sunyaev}}{1973}]{1973AA....24..337S}
{Shakura} N.~I.,  {Sunyaev} R.~A.,  1973, \aap, \href
  {http://adsabs.harvard.edu/abs/1973A%26A....24..337S} {24, 337}

\bibitem[\protect\citeauthoryear{{Smith} et~al.,}{{Smith}
  et~al.}{2002}]{2002AJ....123.2121S}
{Smith} J.~A.,  et~al., 2002, \mn@doi [\aj] {10.1086/339311}, \href
  {http://adsabs.harvard.edu/abs/2002AJ....123.2121S} {123, 2121}

\bibitem[\protect\citeauthoryear{{Stetson}}{{Stetson}}{1987}]{1987PASP...99..191S}
{Stetson} P.~B.,  1987, \mn@doi [\pasp] {10.1086/131977}, \href
  {http://adsabs.harvard.edu/abs/1987PASP...99..191S} {99, 191}

\bibitem[\protect\citeauthoryear{{Suleimanov}, {Meyer}  \&
  {Meyer-Hofmeister}}{{Suleimanov} et~al.}{1999}]{1999AA...350...63S}
{Suleimanov} V.,  {Meyer} F.,   {Meyer-Hofmeister} E.,  1999, \aap, \href
  {http://adsabs.harvard.edu/abs/1999A%26A...350...63S} {350, 63}

\bibitem[\protect\citeauthoryear{{Suleimanov}, {Lipunova}  \&
  {Shakura}}{{Suleimanov} et~al.}{2008}]{2008AA...491..267S}
{Suleimanov} V.~F.,  {Lipunova} G.~V.,   {Shakura} N.~I.,  2008, \mn@doi [\aap]
  {10.1051/0004-6361:200810155}, \href
  {http://adsabs.harvard.edu/abs/2008A%26A...491..267S} {491, 267}

\bibitem[\protect\citeauthoryear{{Thorstensen}, {Charles}  \&
  {Bowyer}}{{Thorstensen} et~al.}{1978}]{1978ApJ...220L.131T}
{Thorstensen} J.,  {Charles} P.,   {Bowyer} S.,  1978, \mn@doi [\apjl]
  {10.1086/182651}, \href {http://adsabs.harvard.edu/abs/1978ApJ...220L.131T}
  {220, L131}

\bibitem[\protect\citeauthoryear{{Troyer} \& {Cackett}}{{Troyer} \&
  {Cackett}}{2016}]{2016arXiv161102578T}
{Troyer} J.~S.,  {Cackett} E.~M.,  2016, preprint, \href
  {http://adsabs.harvard.edu/abs/2016arXiv161102578T} {} (\mn@eprint {arXiv}
  {1611.02578})

\bibitem[\protect\citeauthoryear{{Tudose}, {Fender}, {Linares}, {Maitra}  \&
  {van der Klis}}{{Tudose} et~al.}{2009}]{2009MNRAS.400.2111T}
{Tudose} V.,  {Fender} R.~P.,  {Linares} M.,  {Maitra} D.,   {van der Klis} M.,
   2009, \mn@doi [\mnras] {10.1111/j.1365-2966.2009.15604.x}, \href
  {http://adsabs.harvard.edu/abs/2009MNRAS.400.2111T} {400, 2111}

\bibitem[\protect\citeauthoryear{{Vrtilek}, {Raymond}, {Garcia}, {Verbunt},
  {Hasinger}  \& {Kurster}}{{Vrtilek} et~al.}{1990}]{1990AA...235..162V}
{Vrtilek} S.~D.,  {Raymond} J.~C.,  {Garcia} M.~R.,  {Verbunt} F.,  {Hasinger}
  G.,   {Kurster} M.,  1990, \aap, \href
  {http://adsabs.harvard.edu/abs/1990A%26A...235..162V} {235, 162}

\bibitem[\protect\citeauthoryear{{Welsh}, {Robinson}  \& {Young}}{{Welsh}
  et~al.}{2000}]{2000AJ....120..943W}
{Welsh} W.~F.,  {Robinson} E.~L.,   {Young} P.,  2000, \mn@doi [\aj]
  {10.1086/301486}, \href {http://adsabs.harvard.edu/abs/2000AJ....120..943W}
  {120, 943}

\bibitem[\protect\citeauthoryear{{de Jong}, {van Paradijs}  \&
  {Augusteijn}}{{de Jong} et~al.}{1996}]{1996AA...314..484D}
{de Jong} J.~A.,  {van Paradijs} J.,   {Augusteijn} T.,  1996, \aap, \href
  {http://adsabs.harvard.edu/abs/1996A%26A...314..484D} {314, 484}

\bibitem[\protect\citeauthoryear{{van Paradijs} \& {McClintock}}{{van Paradijs}
  \& {McClintock}}{1994}]{1994AA...290..133V}
{van Paradijs} J.,  {McClintock} J.~E.,  1994, \aap, \href
  {http://adsabs.harvard.edu/abs/1994A%26A...290..133V} {290, 133}

\makeatother
\end{thebibliography}

\bsp	
\label{lastpage}
\end{document}